\documentclass[preprint,showpacs,preprintnumbers,amsmath,amssymb]{revtex4}

\usepackage[pdftex]{graphicx}
\usepackage{dcolumn}
\usepackage{bm}
\usepackage{amsmath}
\usepackage{amssymb}
\usepackage{braket}

\def\lesssim{\ \raise.3ex\hbox{$<$}\kern-0.8em\lower.7ex\hbox{$\sim$}\ }
\def\gesim{\ \raise.3ex\hbox{$>$}\kern-0.8em\lower.7ex\hbox{$\sim$}\ }

\begin{document}
\title{Single-particle properties of a strongly-interacting Bose-Fermi mixture with mass and population imbalance}
\author{Koki Manabe$^1$, Daisuke Inotani$^2$, and Yoji Ohashi$^1$} 
\affiliation{
$^1$ Department of Physics, Keio University, 3-14-1 Hiyoshi, Kohoku-ku, Yokohama 223-8522, Japan \\
$^2$ Research and Education Center for Natural Sciences, Keio University, 4-1-1 Hiyoshi, Kohoku-ku, Yokohama 223-8521, Japan
} 
\begin{abstract}
We theoretically investigate strong-coupling properties of a Bose-Fermi mixture. In the mass- and population-balanced case, two of the authors have shown that a strong hetero-pairing interaction in this mixture brings about coupling phenomena between Fermi atomic excitations and Bose atomic and composite molecular excitations, that appear as an anomalous multiple peak structure in the single-particle spectral weight (SW). [D. Kharga, {\it et. al.}, J. Phys. Soc. Jpn. {\bf 86}, 084301 (2017)]. In this paper, extending this previous work, we show that, although these many-body phenomena are sensitive to mass and population imbalances between the Bose and Fermi components, SW still exhibits the multiple peak structure in a moderately mass-imbalanced $^{87}$Rb-$^{40}$K and $^{23}$Na-$^{40}$K mixtures. We also point out that the photoemission spectrum is a useful quantity to observe this spectral anomaly. Since a real trapped Bose-Fermi mixture is usually accompanied by mass and (local) population imbalance, our results would contribute to the study of a strongly interacting Bose-Fermi mixture, under realistic imbalanced conditions.
\end{abstract}
\maketitle
\par
\section{Introduction}
\par
The high tunability of ultracold atomic gases has contributed to the development of quantum many-body physics discussed in various research fields\cite{Bloch,Giorgini,Georgescu,Gross}: Using an optical lattice technique, Greiner and co-workers have observed the superfluid-Mott insulator transition in a $^{87}$Rb Bose gas\cite{Greiner}. In $^{40}$K\cite{Regal} and $^6$Li Fermi gases\cite{Zwierlein,Kinast,Jochim}, the superfluid phase transition and the BCS (Bardeen-Cooper-Schrieffer)-BEC (Bose-Einstein condensation) crossover phenomenon\cite{Eagles,Leggett,Engelbrecht,BCS-BEC_rev1,BCS-BEC_rev2} have been realized, by using a tunable pairing interaction associated with a Feshbach resonance\cite{Chin}. 
\par
Besides Bose gas and Fermi gas, a gas mixture of Bose and Fermi atoms has also extensively been studied in cold atom physics\cite{dg_KLi1,dg_LiLi1,dg_LiLi2,dg_LiNa1,FR_KRb1,FR_KRb2,FR_KRb3,FR_KRb4,FR_LiCs1,FR_LiNa1,FR_LiNa2,FR_LiRb1,FR_NaK1,FR_NaK2,Mf_KRb1,Mf_KRb2,Mf_KRb3,Mf_NaK1,Mf_NaK2,Mf_NaLi1,dualSF1,dualSF2,dualSF3,inducedint0,cTMA1,cTMA2,cTMA3,cTMA4,cTMA5,cTMA6,iTMA1,iTMA2,others0,others1,others2,Gamma2,Gamma3,Gamma4,stability1,stability2,Bruun1,Bruun2,Bruun3,inducedint1,inducedint3,inducedint4}.  This Bose-Fermi mixture is similar to a $^4$He-$^3$He mixture, as well as quark matter in high-energy physics\cite{QCD1}. Using this similarity, as well as the advantage that one can tune the strength of a Bose-Fermi pairing interaction by using a hetero-nuclear Feshbach resonance, Ref. \cite{QCD1} suggests that this atomic mixture may be used as a quantum simulator for the study of dense QCD matter, where a bound di-quark (boson) and an unpaired quark (fermion) form a nucleon (composite fermion).
\par
In a mass- and population-balanced Bose-Fermi mixture, two of the authors have recently shown that a strong Bose-Fermi pairing interaction causes couplings between Fermi atomic excitations and Bose atomic excitations, as well as atomic excitations and molecular excitations\cite{iTMA1}. As an interesting phenomenon associated with these couplings, the Fermi component of the single-particle spectral weight (SW) has been shown to exhibit a three-peak structure, consisting of two sharp peaks along the free fermion dispersion and composite molecular dispersion, and a broad downward peak being related to Bose single-particle  excitations. Here, we recall that SW in a free Fermi gas only has a single peak line along the free particle dispersion. In a two-component Fermi gas in the BCS-BEC crossover region, SW is known to exhibit a two-peak structure associated with the pseudogap phenomenon originating from strong-pairing fluctuations\cite{FFSW1,FFSW2,FFSW3,Tsuchiya2009}. Thus, the three-peak structure in the Fermi SW is expected to be a characteristic of a strongly interacting Bose-Fermi mixture.
\par
To confirm this expectation, however, one should remember that any real Bose-Fermi mixture is composed of different kinds of atoms or different isotopes, such as $^{87}$Rb-$^{40}$K and $^{23}$Na-$^{40}$K gases, so that it is always accompanied by mass imbalance. In addition, when it is trapped in a harmonic potential, bosons and fermions have different density profiles so that local population imbalance is unavoidable, even when both the components have the same number of atoms. Although a box-type trap has recently been invented\cite{box_Fermi,box_Bose1,box_Bose2} (where a trapped gas is almost uniform), the conventional harmonic trap is still used in many experiments. At this stage, it is unclear to what extent these realistic situations affect the above-mentioned many-body coupling phenomena obtained in the somehow academic mass- and population-balanced case. We also note that highly population-imbalanced Bose-Fermi mixture has recently attracted much attention in the study of Bose polaron\cite{Bose-pol_exp1,Bose-pol_exp2,Bose-pol_theo1,Bose-pol_theo2,Bose-pol_theo3,Bose-pol_theo4,Bose-pol_theo5}.
\par
In this paper, we investigate single-particle properties of a Bose-Fermi mixture with a hetero-nuclear Feshbach resonance. Extending the previous work\cite{iTMA1} to include mass and population imbalances, we examine how these affect strong-coupling corrections to SW. We clarify whether or not the many-body coupling phenomena obtained in the mass- and population-balanced case survive in a mass-imbalanced $^{87}$Rb-$^{40}$K and $^{23}$Na-$^{40}$K mixtures. As an observable quantity related to SW, we also deal with the photoemission spectrum\cite{PES_exp1,PES_exp2,PES_exp3,PES_rev,Ota}.
\par
This paper is organized as follows. In Sec. II, we explain our formulation. We separately examine effects of population imbalance and mass imbalance in Secs. III, and IV, respectively. In Sec. V, we pick up a $^{87}$Rb-$^{40}$K mixture, as well as a $^{23}$Na-$^{40}$K mixture as typical two example of mass-imbalanced Bose-Fermi mixture. Throughout this paper, we set $\hbar=k_{\rm B}=1$, and the system volume $V$ is taken to be unity, for simplicity.
\par
\par
\section{Formulation}
\par
We consider a gas mixture of single-component Bose atoms and single-component Fermi atoms, with a hetero-nuclear Feshbach resonance. This Bose-Fermi mixture is modeled by the Hamiltonian,
\begin{eqnarray}
H=\sum_{{\bm p},{\rm s=B,F}}
\xi^{\rm s}_{\bm p}
c^\dagger_{{\bm p},{\rm s}}
c_{{\bm p},{\rm s}}
-U_{\rm BF}
\sum_{\bm{p,p',q}}
c^\dagger_{\bm{p}+\bm{q},{\rm B}}
c^\dagger_{\bm{p'}-\bm{q},{\rm F}}
c_{{\bm p}',{\rm F}}
c_{{\bm p},{\rm B}},
\label{eq.1}
\end{eqnarray}
where $c^\dagger_{{\bm p},{\rm s}}$ is the creation operator of a Bose (s=B) and a Fermi (s=F) atom. $\xi_{\bm p}^{\rm s}=\varepsilon_{\bm p}^{\rm s}-\mu_{\rm s}={\bm p}^2/(2m_{\rm s})-\mu_{\rm s}$ is the kinetic energy of the s-component, measured from the chemical potential $\mu_{\rm s}$ (where $m_{\rm s}$ is an atomic mass). $-U_{\rm BF}(<0)$ is a Bose-Fermi pairing interaction, which is assumed to be tunable by a hetero-nuclear Feshbach resonance. This contact-type interaction brings about the ultraviolet divergence, which is, as usual, absorbed into the $s$-wave scattering length $a_{\rm BF}$. It is related to the bare pairing interaction $-U_{\rm BF}$ as
\begin{equation}
{4\pi a_{\rm BF} \over m}=
-{U_{\rm BF} \over 1-U_{\rm BF}\sum_{\bm p}^{p_{\rm c}}{m \over {\bm p^2}}}.
\label{eq.2}
\end{equation}
Here, $m=2m_{\rm B}m_{\rm F}/(m_{\rm B}+m_{\rm F})$ is twice the reduced mass, and $p_{\rm c}$ is a cutoff momentum. We measure the interaction strength in terms of the inverse scattering length $(k_{\rm tot}a_{\rm BF})^{-1}$. Here $k_{\rm tot}=(3\pi^2N_{\rm tot})^{1/3}$ is the Fermi momentum in an {\it assumed} two-component Fermi gas with the total number $N_{\rm tot}=N_{\rm F}+N_{\rm B}$ of fermions, where $N_{\rm s=B,F}$ is the particle number in the s-component. In this scale, the interaction strength increases with increasing $(k_{\rm tot}a_{\rm BF})^{-1}$ from the negative value. $(k_{\rm tot}a_{\rm BF})^{-1}=0$ represents the unitarity limit.
\par
This paper only considers a uniform gas, ignoring effects of a harmonic trap, for simplicity. However, the local population imbalance coming from the difference of the density profile between the Bose and the Fermi components in a trap is partially examined by considering the population-imbalanced case ($N_{\rm B}\ne N_{\rm F}$).
\par
Strong-coupling corrections to Bose and Fermi single-particle excitations are conveniently described by the self-energy $\Sigma_{\rm s=B,F}({\bm p},i\omega_{\rm s})$ in the single-particle thermal Green's functions,
\begin{align}
G_{\rm s=B,F}(\bm{p},i\omega_{\rm s})=\frac{1}{i\omega_{\rm s}-\xi_{\bm{p}}^{\rm s}-\Sigma_{\rm s}(\bm{p},i\omega_{\rm s})},
\label{eq.3}
\end{align}
where $i\omega_{\rm B}$ ($i\omega_{\rm F}$) is the boson (fermion) Matsubara frequency.
\par
\begin{figure}
\includegraphics[width=10cm]{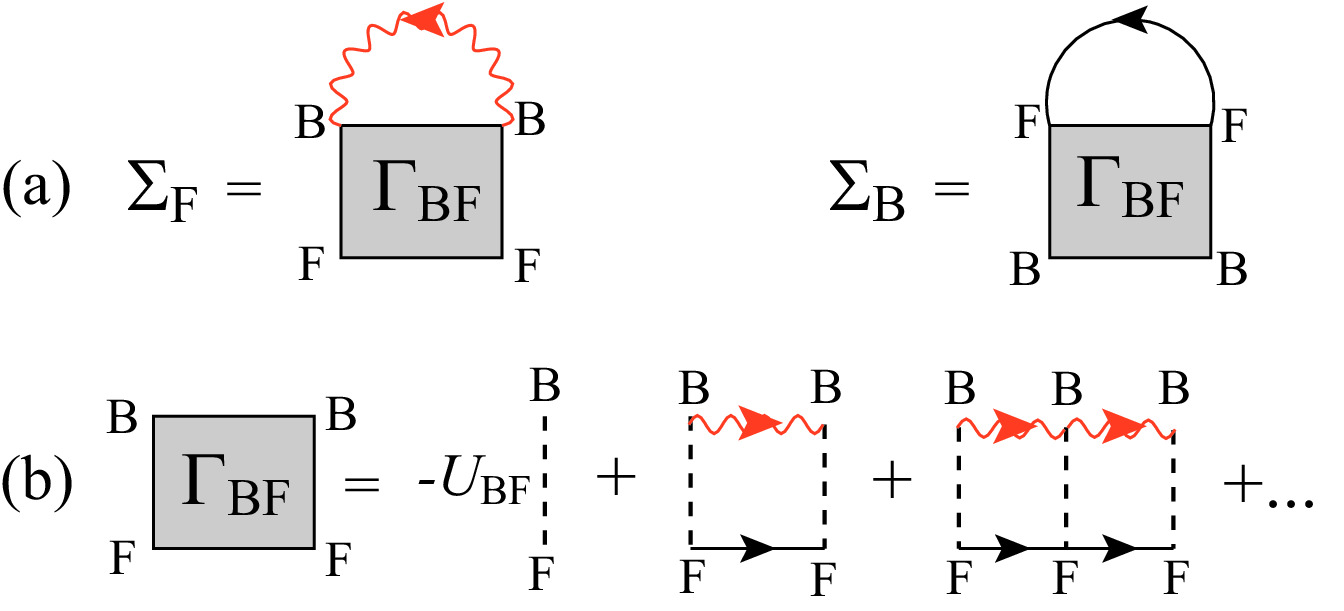}
\caption{(a) Self-energies $\Sigma_{\rm s=B,F}$ in iTMA. (b) Bose-Fermi scattering matrix $\Gamma_{\rm BF}({\bm q},i\omega_{\rm F})$ in Eq. (\ref{eq.10}), physically describing hetero-pairing fluctuations. The solid line is the bare Fermi Green's function $G_{\rm F}^0$ in Eq. (\ref{eq.9}). The wavy line is the {\it modified} Bose Green's function $\tilde{G}_{\rm B}^0$ in Eq.\eqref{eq.5}. The dashed line is the hetero-pairing interaction $-U_{\rm BF}$.
}
\label{fig1}
\end{figure}
\par
We evaluate $\Sigma_{\rm s}({\bm p},i\omega_{\rm s})$ within the framework of the $T$-matrix-type approximation developed in Ref. \cite{iTMA1}, which is diagrammatically described in Fig.\ref{fig1}. This diagrammatic structure is formally the same as that in the ordinary (non-selfconsistent) $T$-matrix approximation (TMA); however, the crucial difference is that the bare Bose Green's function,
\begin{equation}
G_{\rm B}^0(\bm{p},i\omega_{\rm B})=
{1 \over i\omega_{\rm B}-\xi_{\bm{p}}^{\rm B}},
\label{eq.4}
\end{equation}
in the ordinary TMA self-energy diagrams is now replaced by the {\it modified} one\cite{iTMA1},
\begin{align}
\tilde{G}_{\rm B}^0(\bm{p},i\omega_{\rm B})
=\frac{1}{i\omega_{\rm B}-\tilde{\xi}_{\bm{p}}^{\rm B}}.
\label{eq.5}
\end{align}
Here, $\tilde{\xi}_{\bm{p}}^{\rm B}=\xi_{\bm{p}}^{\rm B}+\Sigma_{\rm B}(\bm{0},0)\equiv \varepsilon_{\bm{p}}^{\rm B}-\tilde{\mu}_{\rm B}$ (where ${\tilde \mu}_{\rm B}=\mu_{\rm B}-\Sigma_{\rm B}(0,0)$) involves the self-energy correction at $\bm{p}=i\omega_{\rm B}=0$. This is motivated by the Hugenholtz-Pines theorem\cite{HP},
\begin{align}
\mu_{\rm B}-\Sigma_{\rm B}(\bm{p}=\bm{0},i\omega_{\rm B}=0)=0,
\label{eq.6}
\end{align}
stating that the Bose excitations become gapless at the BEC phase transition. The modified Green's function ${\tilde G}_{\rm B}^0({\bm p},i\omega_{\rm B})$ in Eq. (\ref{eq.5}) is chosen so as to satisfy this required condition at $T_{\rm c}$. We briefly note that the bare Bose Green's function $G_{\rm B}^0$ in Eq. (\ref{eq.4}) still has a {\it gapped} Bose excitation spectrum at $T_{\rm c}$, which means that the self-energy in TMA underestimates effects of low-energy Bose excitations near $T_{\rm c}$.
\par
In this {\it improved} $T$-matrix approximation (iTMA), the summation of the diagrams in Fig. \ref{fig1} gives
\begin{align}
\Sigma_{\rm B}(\bm{p},i\omega_{\rm B})&= T\sum_{\bm{q},i\omega_{\rm F}'}\Gamma_{\rm BF}(\bm{q},i\omega_{\rm F}')G_{\rm F}^0(\bm{q}-\bm{p},i\omega_{\rm F}'-i\omega_{\rm B}),
\label{eq.7}
\\
\Sigma_{\rm F}(\bm{p},i\omega_{\rm F})&=-T\sum_{\bm{q},i\omega_{\rm F}'}\Gamma_{\rm BF}(\bm{q},i\omega_{\rm F}')\tilde{G}_{\rm B}^0(\bm{q}-\bm{p},i\omega_{\rm F}'-i\omega_{\rm F}),
\label{eq.8}
\end{align}
where
\begin{equation}
 G_{\rm F}^0(\bm{p},i\omega_{\rm F})
 ={1 \over i\omega_{\rm F}-\xi_{\bm{p}}^{\rm F}}
\label{eq.9}
\end{equation}
is the bare Fermi single-particle Green's function, and 
\begin{eqnarray}
\Gamma_{\rm BF}(\bm{q},i\omega_{\rm F})
&=&
-{U_{\rm BF} \over 1-U_{\rm BF}\Pi_{\rm BF}({\bm q},i\omega_{\rm F})}
\nonumber
\\
&=&
\frac{1}{\frac{m}{4\pi a_{\rm BF}}+\Bigl[\Pi_{\rm BF}({\bm q},i\omega_{\rm F})-\sum_{\bm{k}}^{p_{\rm c}}\frac{m}{k^2}\Bigr]}
\label{eq.10}
\end{eqnarray}
is the iTMA Bose-Fermi scattering matrix, physically describing hetero-pairing fluctuations. Here, 
\begin{align}
\Pi_{\rm BF}(\bm{q},i\omega_{\rm F})
&=-T\sum_{\bm{p},i\omega_{\rm B}}G_{\rm F}^0(\bm{q}-\bm{p},i\omega_{\rm F}-i\omega_{\rm B})\tilde{G}_{\rm B}^0(\bm{p},i\omega_{\rm B})
\nonumber
\\
&=\sum_{\bm{p}}\frac{1-f(\xi_{\bm{q}-\bm{p}}^{\rm F})+n_{\rm B}(\tilde{\xi}_{\bm{p}}^{\rm B})}{\xi_{\bm{q}-\bm{p}}^{\rm F}+\tilde{\xi}_{\bm{p}}^{\rm B}-i\omega_{\rm F}}
\label{eq.11}
\end{align}
is the hetero-pair correlation function, where $n_{\rm B}(x)$ and $f(x)$ are the Bose and Fermi distribution function, respectively.
\par
The BEC phase transition temperature $T_{\rm c}$ is conveniently determined from the Hugenholtz-Pines condition in Eq. (\ref{eq.6}). We actually solve this equation, together with the equation for the number $N_{\rm B}$ ($N_{\rm F}$) of Bose (Fermi) atoms,
\begin{align}
N_{\rm B}&=-T\sum_{\bm{p},i\omega_{\rm B}}G_{\rm B}(\bm{p},i\omega_{\rm B}),
\label{eq.12}
\\
N_{\rm F}&= T\sum_{\bm{p},i\omega_{\rm F}}G_{\rm F}(\bm{p},i\omega_{\rm F}),
\label{eq.13}
\end{align}
to self-consistently determine $T_{\rm c}$, $\mu_{\rm B}(T_{\rm c})$, and $\mu_{\rm F}(T_{\rm c})$. Above $T_{\rm c}$, we only deal with the number equations (\ref{eq.12}) and (\ref{eq.13}), to evaluate $\mu_{\rm B}(T)$, and $\mu_{\rm F}(T)$.
\par
The single-particle spectral weights (SWs) $A_{s={\rm B,F}}(\bm{p},\omega)$ are related to the analytic-continued Green's functions as,
\begin{align}
A_{\rm s=B,F}(\bm{p},\omega)&=-\frac{1}{\pi}{\rm Im}
\left[G_{\rm s}(\bm{p},i\omega_{\rm s}\to \omega+i\delta\equiv\omega_+)
\right],
\label{eq.14}
\end{align}
where $\delta$ is an infinitesimally small positive number. The photoemission spectra (PESs) $I_{\rm s=B,F}({\bm p},\omega)$\cite{PES_exp1,PES_exp2,PES_exp3} are then immediately obtained from SWs as\cite{PES_rev,Ota}, under the assumption that the final state interaction is absent,
\begin{align}
I_{\rm F}(\bm{p},\omega)=2\pi t_{\rm F}^2p^2A_{\rm F}(\bm{p},\omega)f(\omega),
\label{eq.15}
\\
I_{\rm B}(\bm{p},\omega)=2\pi t_{\rm B}^2p^2A_{\rm B}(\bm{p},\omega)n_{\rm B}(\omega).
\label{eq.16}
\end{align}
Here, $t_{\rm s}$ is a transfer-matrix element from the initial atomic hyperfine state $|I\rangle$ to the final one $|F\rangle$ ($\ne |I\rangle$). Between the two, the Fermi SW will be found to be more useful for the study of many-body coupling phenomena mentioned in Sec. I. Thus, we only examine the corresponding Fermi PES in this paper.
\par
\begin{figure}
\includegraphics[width=8.5cm]{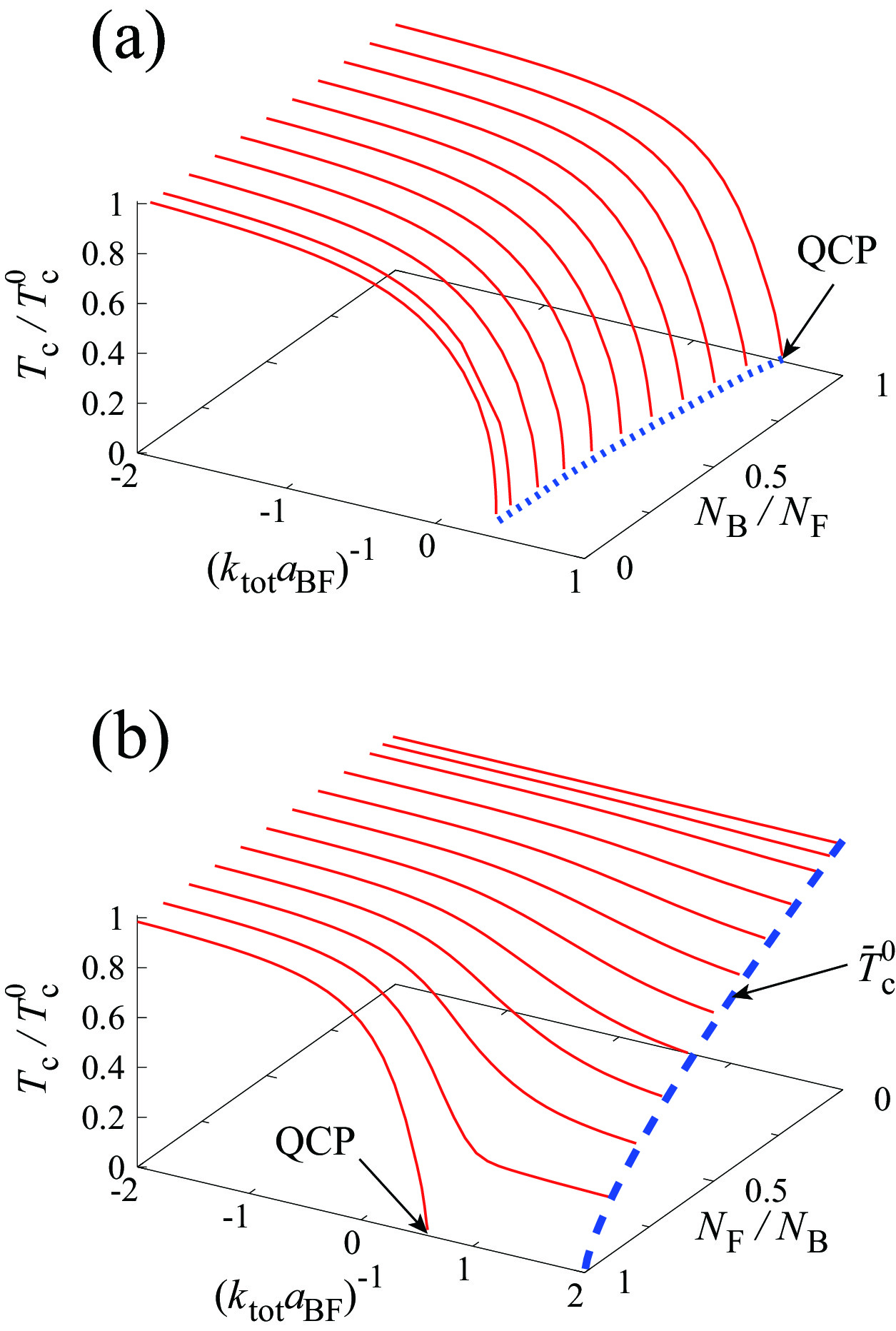}
\caption{Calculated Bose-Einstein condensation temperature $T_c$ in a Bose-Fermi mixture with population imbalance ($N_{\rm B}\ne N_{\rm F}$). We set $m_{\rm B}=m_{\rm F}$. (a) $N_{\rm B}<N_{\rm F}$. (b) $N_{\rm B}>N_{\rm F}$. The dashed line shows ${\bar T}_{\rm c}^0$ in Eq. (\ref{eq.3.2}).  $T_{\rm c}^0$ is the BEC phase transition temperature in an ideal gas of $N_{\rm B}$ bosons in Eq (\ref{eq.3.1}). `QCP' is the quantum critical point at which $T_{\rm c}$ vanishes.
}
\label{fig2}
\end{figure}
\par
\section{Single-particle excitations in a population-imbalanced Bose-Fermi mixture}
Figure \ref{fig2} shows the BEC transition temperature $T_{\it c}$ in a Bose-Fermi mixture with population imbalance ($N_{\rm B}\neq N_{\rm F}$). In the weak-coupling limit $(k_{\rm tot}a_{\rm BF})^{-1}\ll -1$, since Fermi atoms do not affect the BEC phase transition, $T_{\rm c}$ is simply given by the BEC phase transition temperature in an ideal Bose gas,
\begin{equation}
T_{\rm c}^0={2\pi \over m_{\rm B}}
\left(
{N_{\rm B} \over \zeta(3/2)}
\right)^{2/3},
\label{eq.3.1}
\end{equation}
where $\zeta(3/2)\simeq 2.613$ is the zeta function. Starting from this extreme case, when $N_{\rm B}/N_{\rm F}<1$, we see in Fig. \ref{fig2}(a) that the overall interaction dependence of $T_{\rm c}$ is similar to that in the population-balanced case ($N_{\rm B}=N_{\rm F}$)\cite{iTMA1}: $T_{\rm c}$ gradually decreases from $T_{\rm c}^0$ with increasing the interaction strength, to eventually vanish around the unitarity limit $(k_{\rm tot}a_{\rm BF})^{-1}=0$. This vanishing $T_{\rm c}$ is because of the formation of Bose-Fermi molecules in the two-body level when $(k_{\rm tot}a_{\rm BF})^{-1}>0$, and the most Bose atoms pair up with Fermi atoms to become composite molecules there. As a result, BEC of unpaired {\it Bose atoms} no longer occurs, when the interaction strength exceeds a quantum critical point (QCP) at $(k_{\rm tot}a_{\rm BF})^{-1}\sim 0$, as seen in Fig. \ref{fig2}(a).
\par
When $N_{\rm B}>N_{\rm F}$ (Fig. \ref{fig2}(b)), on the other hand, the BEC phase transition remains to exist ($T_{\rm c}>0)$ even in the strong-coupling limit ($(k_{\rm tot}a_{\rm BF})^{-1}\gg +1$). This is simply because the number $\Delta N_{\rm B}\equiv N_{\rm B}-N_{\rm F}(>0)$ of bosons remain unpaired in this limit. Indeed, $T_{\rm c}$ in the strong-coupling limit is well described by their BEC transition temperature ${\bar T}_{\rm c}^0$,
\begin{equation}
{\bar T}_{\rm c}^0={2\pi \over m_{\rm B}}
\left(
{\Delta N_{\rm B} \over \zeta(3/2)}
\right)^{2/3},
\label{eq.3.2}
\end{equation}
as shown in Fig.\ref{fig2}(b).
\par
\begin{figure}
\includegraphics[width=8.5cm]{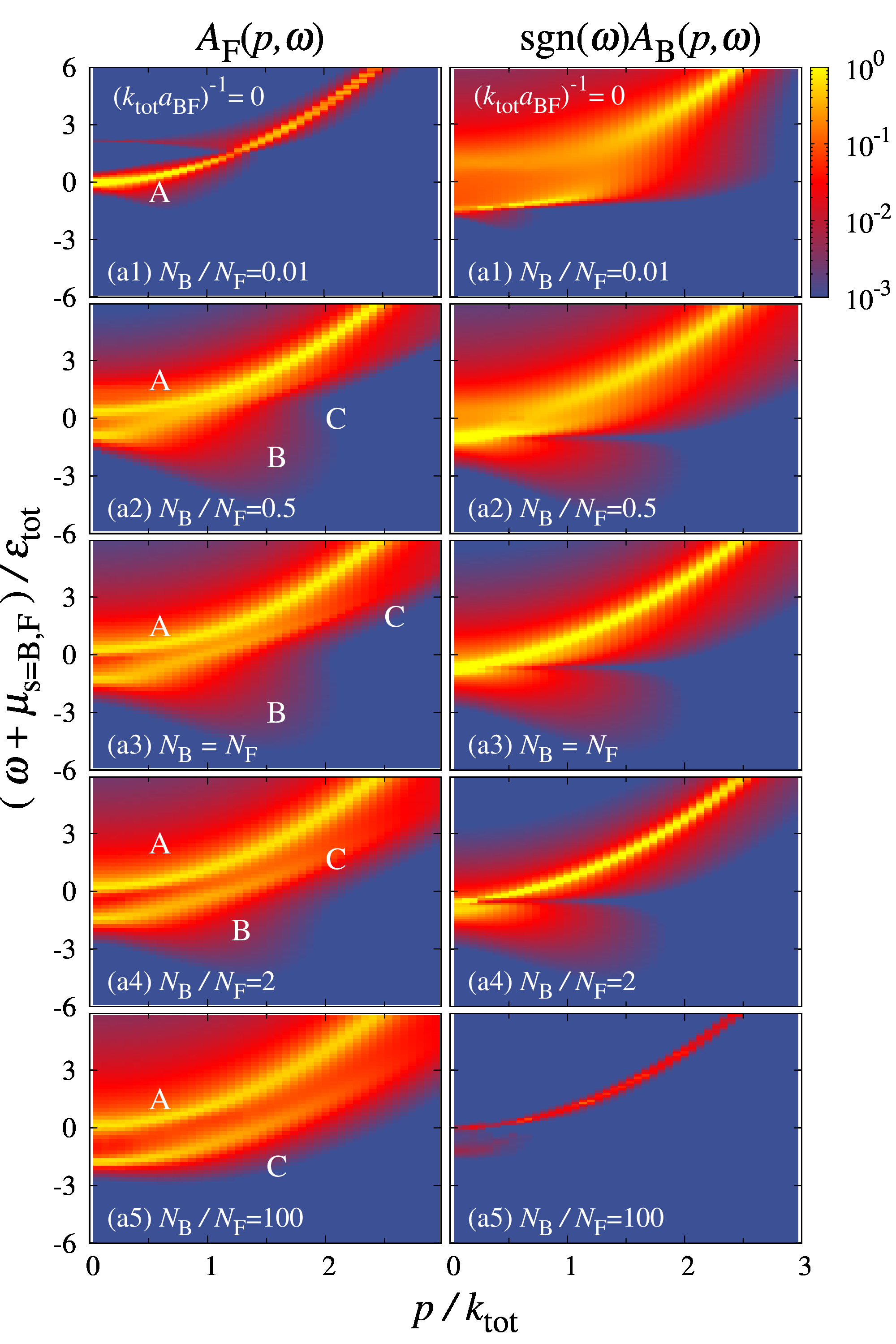}
\caption{Calculated intensity of single-particle spectral weights $A_{{\rm s}={\rm B,F}}(\bm{p},\omega)$ in a {\it population-imbalanced} unitary Bose-Fermi mixture at $T=T_{\rm c}$. (a1)-(a5) $A_{\rm F}(\bm{p},\omega)$. (b1)-(b5) $A_{\rm B}(\bm{p},\omega)\times{\rm sgn}(\omega)$. We set $m_{\rm B}=m_{\rm F}$. The spectral intensity is normalized by $\varepsilon_{\rm tot}^{-1}\equiv 2m/k_{\rm tot}^2$.
}
\label{fig3}
\end{figure}
\par
Figures \ref{fig3}(a1)-(a5) show the Fermi SW $A_{\rm F}(\bm{p},\omega)$ in the unitarity limit at $T_{\rm c}$. In the population-balanced case (panel (a3)), the Fermi SW exhibits a three-peak structure as a result of the Fermi-Bose and atom-molecule couplings\cite{iTMA1} mentioned in Sec. I: (A) Sharp peak line along the Fermi free-particle dispersion, $\omega=\xi_{\bm{p}}^{\rm F}$. (B) Broad downward peak in the negative energy region around $\omega=-\tilde{\xi}_{\bm p}^{\rm B}$, where the Bose dispersion ${\tilde \xi}_{\bm p}^{\rm B}$ is given below Eq. (\ref{eq.5}). (C) Sharp upward peak line along the composite molecular dispersion, $\omega=\xi_{\bm p}^{\rm CF}\equiv {\bm p}^2/(2M_{\rm CF})-\mu_{\rm CF}$, where $M_{\rm CF}\simeq m_{\rm B}+m_{\rm F}$ is a molecular mass and $\mu_{\rm CF}$ is the molecular chemical potential\cite{comment4}. 
\par
When the ratio $N_{\rm B}/N_{\rm F}$ decreases from unity (Fig.s \ref{fig3}(a3)$\to$(a1)), the two peaks (B) and (C) are found to gradually disappear. This is simply because the system approaches a single-component free Fermi gas. In the extreme population-imbalanced case shown in Fig. \ref{fig3}(a1), the peak line (A) is only seen, as expected. 
\par
In the opposite case, on the other hand, with increasing the ratio $N_{\rm B}/N_{\rm F}>1$, while the broad peak (B) gradually disappears, the molecular peak (C) continues to exist, in addition to the free fermion dispersion (A), even in Fig. \ref{fig3}(a5).
\par
To understand these population-imbalance effects on the Fermi SW, we conveniently approximate the iTMA Fermi Green's function $G_{\rm F}({\bm p},i\omega_{\rm F})$ to\cite{iTMA1} 
\begin{equation}
G_{\rm F}(\bm{p},i\omega_{\rm F}\rightarrow\omega_+)\simeq
{1 \over 
\displaystyle
\omega_+-\xi_{\bm{p}}^{\rm F}-\frac{ZN_{\rm B}^0}{\omega_+-\xi_{\bm{p}}^{\rm CF}}-\left\langle\frac{ZN_{\rm CF}}{\omega_++\tilde{\xi}_{{\bm{k}}_{\rm CF}-\bm{p}}^{\rm B}}\right\rangle_{{\bm k}_{\rm CF}}
}.
\label{eq.3.3}
\end{equation}
(For the outline of the derivation, see the Appendix.) In Eq. (\ref{eq.3.3}), $N_{\rm CF}=\sum_{\bm{q}}f(\xi_{\bm{q}}^{\rm CF})$ is the number of composite Fermi molecules with the chemical potential $\mu_{\rm CF}>0$, $N_{\rm B}^0=\sum_{\bm{q}}n(\tilde{\xi}_{\bm{q}}^{\rm B})$, and $|{\bm k}_{\rm CF}|=\sqrt{2M_{\rm CF}\mu_{\rm CF}}$ gives the size of the Fermi surface in the composite Fermi molecular gas. The average $\langle\cdots\rangle_{\bm Q}$ is taken over the direction of ${\bm Q}$, and $Z$ is a positive constant. (For details of $Z$, we refer to Ref. \cite{iTMA1}.) Equation (\ref{eq.3.3}) explains that hetero-pairing fluctuations couple the Fermi atomic excitations $\omega=\xi_{\rm p}^{\rm F}$ (A) with the molecular excitations $\omega=\xi_{\rm p}^{\rm CF}$ (C) with the coupling strength $ZN_{\rm B}^0$, as well as with the Bose excitations $\omega=-{\tilde \xi}^{\rm B}_{{\bm k}_{\rm CF}-{\bm p}}$ (B) with the coupling strength $ZN_{\rm CF}$, respectively. For the latter Fermi-Bose coupling, because of the average over the direction of ${\bm k}_{\rm CF}$ in Eq. (\ref{eq.3.3}), the Bose excitations gives the broad spectrum structure in the negative energy region of Fig. \ref{fig3}(a3). 
\par
As the number $N_{\rm B}$ of Bose atoms decreases ($N_{\rm B}/N_{\rm F}<1$), both $N_{\rm B}^0$ and $N_{\rm CF}$ in the denominator in Eq. (\ref{eq.3.3}) decrease, to eventually vanish in the limit $N_{\rm B}\to 0$. This immediately explains the single-peak structure in the Fermi SW in Fig. \ref{fig3}(a1). 
\par
On the other hand, when $N_{\rm F}$ decreases ($N_{\rm B}/N_{\rm F}>1$), while $N_{\rm CF}$ vanishes in the large population-imbalance limit, $N_{\rm B}^0$ would approach the non-zero value $N_{\rm B}$. Thus, Eq. (\ref{eq.3.3}) is reduced to 
\begin{equation}
G_{\rm F}(\bm{p},\omega_+)=
{1 \over \displaystyle 
\omega_+-\xi_{\bm{p}}^{\rm F}-\frac{ZN_{\rm B}^0}{\omega_+-\xi_{\bm{p}}^{\rm CF}}},
\label{eq.3.4}
\end{equation}
which has the two poles,
\begin{equation}
E_{\bm p}^\pm=
{1 \over 2}
\left[
[\xi_{\bm p}^{\rm F}+\xi_{\bm p}^{\rm CF}]
\pm
\sqrt{[\xi_{\bm p}^{\rm F}-\xi_{\bm p}^{\rm CF}]^2+4ZN_{\rm B}^0}
\right].
\label{eq.3.5}
\end{equation}
Equation (\ref{eq.3.5}) explains the two-peak structure in Fig.\ref{fig3}(a5). That is, the Fermi single-particle excitations in the highly population-imbalanced regime ($N_{\rm B}/N_{\rm F}\gg 1$) are dominated by the atom-molecule coupling phenomenon. We briefly note that this limit just corresponds to the Bose-polaron system.
\par
Applying the same approximation to the Bose component, we obtain\cite{iTMA1},
\begin{equation}
G_{\rm B}(\bm{p},i\omega_{\rm F}\to\omega_+)\simeq
{1 \over 
\displaystyle
\omega_+-\xi_{\bm p}^{\rm B}
-
\left\langle
{ZN_{\rm F}^0
\over
\omega_+-\xi_{{\tilde {\bm k}}_{\rm F}-\bm{p}}^{\rm CF}
}
\right\rangle_{{\tilde {\bm k}}_{\rm F}}
-
\left\langle
{
ZN_{\rm CF}
\over 
\omega_++{\xi_{{\bm k}_{\rm CF}-\bm{p}}^{\rm F}}
}
\right\rangle_{{\tilde {\bm k}}_{\rm CF}}
}.
\label{eq.3.6}
\end{equation}
Here, $N_{\rm F}^0=\sum_{\bm{q}}f(\xi_{\bm{q}}^{\rm F})$, and $|\tilde{\bm{k}}_{\rm F}|=\sqrt{2m_{\rm F}\mu_{\rm F}}$. Equation (\ref{eq.3.6}) shows that the Bose single-particle excitations ($\omega=\xi_{\bm p}^{\rm B}$) couple with composite Fermi molecular excitations ($\omega=\xi_{{\tilde k}_{\rm F}-{\bm p}}^{\rm CF}$), as well as Fermi hole excitations ($\omega=-\xi^{\rm F}_{{\bm k}_{\rm CF}-{\bm p}}$); however, because of the angular averages in the denominator of Eq. (\ref{eq.3.6}), the three-peak structure is not clearly seen in the Bose SW $A_{\rm B}({\bm p},\omega)$, when $N_{\rm B}=N_{\rm F}$ (see Fig. \ref{fig3}(b3)). 
\par
When the number $N_{\rm F}$ of Fermi atoms decreases ($N_{\rm B}/N_{\rm F}>1$), both $N_{\rm F}^0$ and $N_{\rm CF}$ decrease. Thus, the Bose SW is gradually reduced to the spectral weight in a free Bose gas (where the single peak line is along $\omega=\xi_{\bm p}^{\rm B}$), as seen in Figs. \ref{fig3}(b3)$\to$(b5). 
\par
With increasing $N_{\rm F}$ ($N_{\rm B}/N_{\rm F}<1$), the system eventually reaches the situation that $N_{\rm F}^0\to N_{\rm F}$ and $N_{\rm CF}\to 0$. Because of this, the spectral structure in Figs. \ref{fig3}(b1) and (b2) are dominated by Bose atomic excitations and broad composite molecular excitations, but the downward broad peak associated with Fermi hole excitations becomes weak.
\par
The above discussions indicate that the Fermi SW $A_{\rm F}({\bm p},\omega)$ is more suitable than the Bose SW $A_{\rm B}({\bm p},\omega)$, for the study of the Fermi-Bose and atom-molecule coupling phenomena. 
\par
\begin{figure}
\includegraphics[width=10cm]{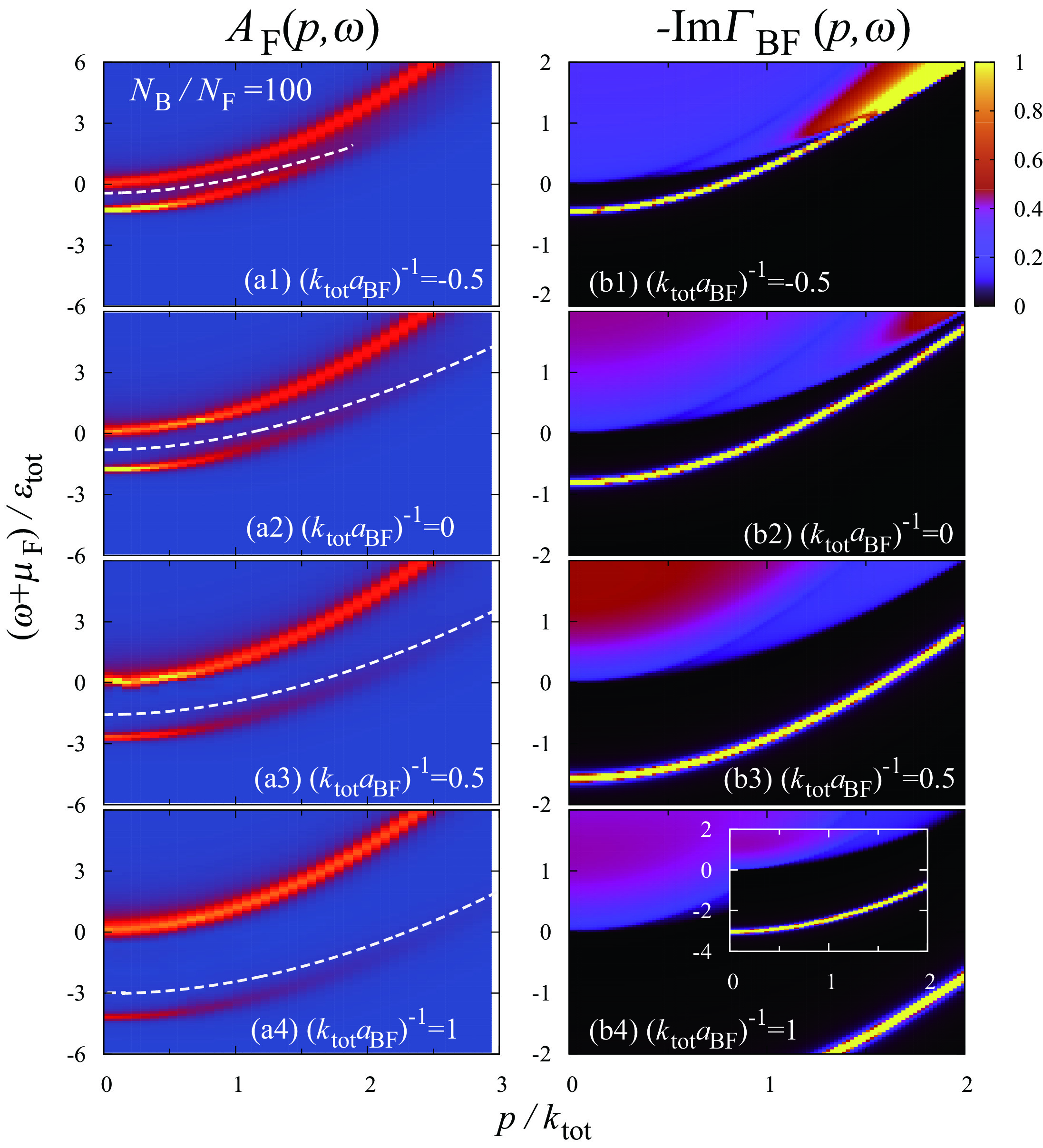}
\caption{(a1)-(a4) Calculated intensity of the Fermi SW $A_{\rm F}(\bm{p},\omega)$ in the highly population imbalanced Bose-Fermi mixture ($N_{\rm B}/N_{\rm F}=100\gg 1$) at $T_{\rm c}$. The intensity is normalized by $\varepsilon_{\rm tot}^{-1}$. (b1)-(b4) Corresponding spectrum of the Bose-Fermi scattering matrix $-{\rm Im}[\Gamma_{\rm BF}(q,i\omega_{\rm F}\to\omega_+)]$, normalized by $2\pi^2/(mk_{\rm tot})$. The inset in panel (b4) shows the spectrum in the wide energy region $-4\le (\omega+\mu_{\rm F})/\varepsilon_{\rm tot}\le 2$, to clearly show the sharp peak line associated with molecular excitations. In each left panel, the dashed line shows the pole position of $\Gamma_{\rm BF}({\bm p},i\omega_{\rm F}\to\omega_+)$.
}
\label{fig4}
\end{figure}
\par
The atom-molecule coupling phenomenon seen in Figs. \ref{fig3}(a4) and (a5) also appears away from the unitarity limit, as shown in the left panels in Fig. \ref{fig4}. In this figure, when $(k_{\rm tot}a_{\rm BF})^{-1}>0$, the appearance of the lower sharp peak associated with molecular excitations would be reasonable, because the pairing interaction is strong enough to produce two-body bound states there. On the other hand, Fig. \ref{fig4}(a1) shows that the Fermi SW still has a sharp molecular peak line, in spite of the absence of a two-body bound state when $(k_{\rm tot}a_{\rm BF})^{-1}=-0.5<0$. 
\par
Regarding this, plotting the spectrum ${\rm Im}[\Gamma_{\rm BF}({\bm p},i\omega_{\rm F}\to\omega_+)]$ of the Bose-Fermi scattering matrix, one finds an isolated sharp peak line (which physically describes composite Fermi molecular excitations) below the continuum spectrum, not only in the strong-coupling regime (Figs. \ref{fig4}(b2)-(b4)), but also in the weak-coupling region (Fig. \ref{fig4}(b1)). The latter result implies the stabilization of a Bose-Fermi bound state by a many-body (medium) effect. To see this in a simple manner, we approximate Eq. (\ref{eq.10}) at $T_{\rm c}$ to, after taking the analytic continuation $i\omega_{\rm F}\to\omega_+$, 
\begin{equation}
\Gamma_{\rm BF}(\bm{q},\omega_+)\simeq
{1 \over \displaystyle
\frac{m}{4\pi a_{\rm BF}}
+
\sum_{\bm{p}}\left[\frac{1}{\varepsilon^{\rm F}_{\bm{q}-\bm{p}}+\varepsilon^{\rm B}_{\bm{p}}-{\tilde \omega}_+}-\frac{m}{p^2}
\right]
-N_{\rm B}^0G_{\rm F}^0(\bm{q},\omega_+)},
\label{eq.3.7}
\end{equation}
where ${\tilde \omega}=\omega+\mu_{\rm F}$, and we have used the fact that the Bose distribution function $n_{\rm B}({\tilde \xi}^{\rm B}_{\bm p})$ in Eq. (\ref{eq.11}) diverges at ${\bm p}=0$ at $T_{\rm c}$. When one ignores the last term in the denominator in Eq. (\ref{eq.3.7}), the pole equation of Eq. (\ref{eq.3.7}) is essentially the same as the two-body bound-state equation (which has a solution only when $a_{\rm BF}>0$). The term $N_{\rm B}^0G_{\rm F}({\bm q},\omega_+)$ in Eq. (\ref{eq.3.7}) physically describes medium effects. Including this at ${\bm q}=0$, we obtain the pole equation, 
\begin{equation}
1=a_{\rm BF}
\left[
\sqrt{m|{\tilde \omega}|}-{4\pi N_{\rm B}^0 \over m|{\tilde \omega}|}
\right].
\label{eq.3.8}
\end{equation}
This indicates that a Bose-Fermi bound state is also possible, when $(k_{\rm tot}a_{\rm BF})^{-1}\le 0$ (where the two-body bound state is absent) in the many-particle case\cite{Gamma2,Gamma3,Gamma4}. At the unitarity ($a_{\rm BF}^{-1}=0$), for example, Eq. (\ref{eq.3.8}) gives
\begin{equation}
{\tilde \omega}=-{(4\pi N_{\rm B}^0)^{2 \over 3} \over m}.
\label{eq.3.9}
\end{equation}
In Fig. \ref{fig4}, the lower peak line in each left panel is found to be close to the bound-state dispersion (dashed line) in the right panel, although the former is somehow pushed down, due to the coupling with the Fermi atomic excitations $\omega=\xi_{\bm p}^{\rm F}$ (see Eq. (\ref{eq.3.5})). In this figure, with increasing the strength of a hetero-pairing interaction (panel (a1)$\to$(a4)), the character of a Bose-Fermi molecule continuously changes from the many-body bound state assisted by medium, to the two-body bound state. We briefly note that a similar crossover phenomenon from a polaron state to the two-body bound state (polaron-molecule crossover) has been discussed in the Bose-polaron system at $T=0$\cite{Bose-pol_theo2}.
\par
\begin{figure}
\includegraphics[width=10cm]{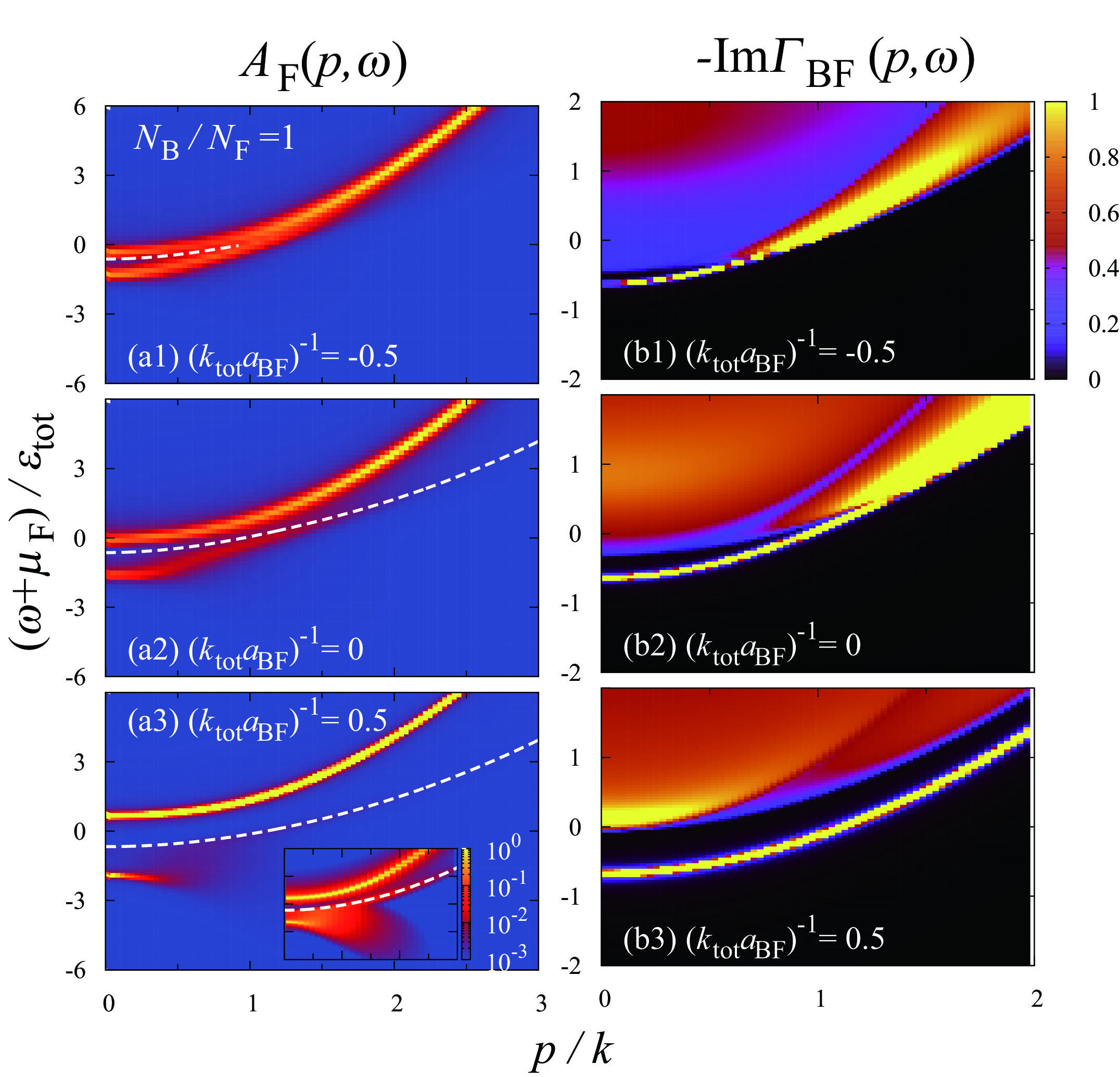}
\caption{Same plot as Fig.\ref{fig4} for the population-balanced case $N_{\rm B}=N_{\rm F}$. The inset in panel (a3) shows the logarithmic plot to show the three-peak structure discussed in Fig. \ref{fig3}.
\label{fig5}
}
\end{figure}

The above discussions are also applicable to the population-balanced case ($N_{\rm B}=N_{\rm F}$): In Fig. \ref{fig5}, the spectrum $-{\rm Im}[\Gamma_{\rm BF}({\bm p},i\omega_{\rm F}\to\omega_+)]$ of the Bose-Fermi scattering matrix (right panels) has an isolated sharp peak, bringing out the lower peak in the Fermi SW $A_{\rm F}({\bm p},\omega)$ (left panels). In the right panels, the peak energy is lowered as the interaction strength increases, because of the increase of the binding energy of a Bose-Fermi bound state. This tendency is the same as the highly population-imbalanced case shown in Fig. \ref{fig4}.
\par
\par
\begin{figure}
\includegraphics[width=16cm]{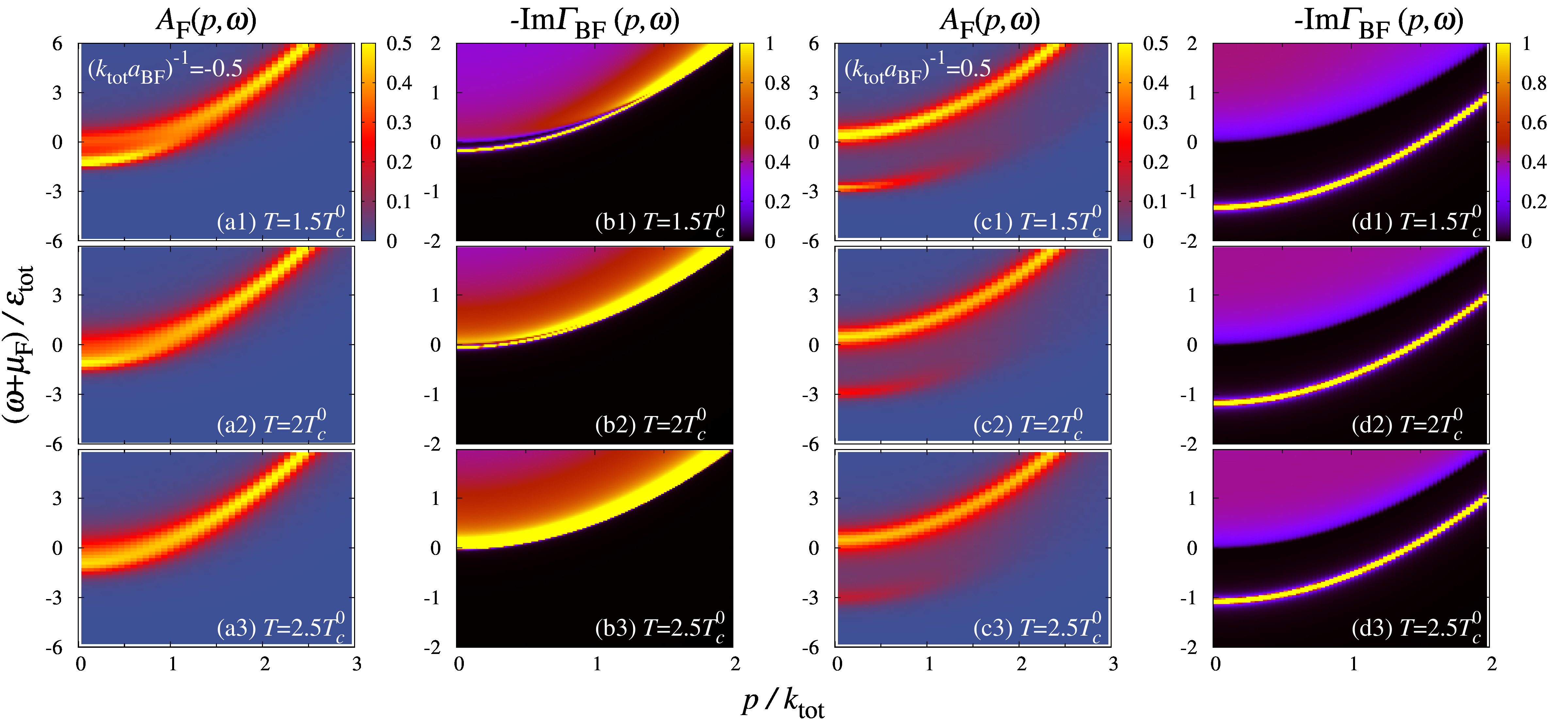}
\caption{Intensity of the Fermi SW $A_{\rm F}(\bm{p},\omega)$, as well as the spectrum $-{\rm Im}[\Gamma_{\rm BF}(q,i\omega_{\rm F}\rightarrow\omega_+)]$ of the Bose-Fermi scattering matrix, in a highly population-imbalanced Bose-Fermi mixture ($N_{\rm B}/N_{\rm F}=100$). The normalization of the spectral intensity is the same as that in Fig. \ref{fig4}. (a1)-(a3) and (b1)-(b3): Weak-coupling case $(k_{\rm tot}a_{\rm BF})^{-1}=-0.5$. (c1)-(c3) and (d1)-(d3): Strong-coupling case $(k_{\rm tot}a_{\rm BF})^{-1}=0.5$.
}
\label{fig6}
\end{figure}
\par
In the normal state above $T_{\rm c}$, we expect the following two thermal effects: (1) The Bose distribution function $n_{\rm B}({\tilde \xi}^{\rm B}_{\bm q})$ no longer diverges at ${\bm q}=0$ when $T>T_{\rm c}$, so that the approximation giving the first term in the last line in Eq. (\ref{eq.A2}) becomes worse. Roughly speaking, this would lead to the broadening of the peak line coming from molecular excitations in the Fermi SW $A_{\rm F}({\bm p},\omega)$. In addition, because the factor $ZN_{\rm B}^0=Z\sum_{\bm q}n_{\rm B}({\tilde \xi}^{\rm B}_{\bm q})$ decreases with increasing the temperature, the atom-molecule coupling also becomes weak. (2) When Bose-Fermi bound states thermally dissociate into unpaired atoms at high temperatures, the approximate expression for the Bose-Fermi scattering matrix $\Gamma_{\rm BF}({\bm q},i\omega_{\rm F})$ in Eq. (\ref{eq.A1}) is no longer valid. 
\par
Keeping these two thermal effects in mind, we find in Figs. \ref{fig6}(a1)-(a3) ($(k_{\rm tot}a_{\rm BF})^{-1}=-0.5<0$) that the molecular peak line soon becomes obscure with increasing the temperature above $T_{\rm c}$. In this case, because of the weak Bose-Fermi pairing interaction, the sharp spectral peak in $-{\rm Im}[\Gamma_{\rm BF}(q,i\omega_{\rm F}\rightarrow\omega_+)]$ describing molecular excitations also soon disappears above $T_{\rm c}$ (see Figs. \ref{fig6}(b1)-(b3)). Thus, the above-mentioned two thermal effects are considered to suppress the atom-molecular coupling in the Fermi SW $A_{\rm F}({\bm p},\omega)$ in the weak-coupling case.
\par
When $(k_{\rm tot}a_{\rm BF})^{-1}=0.5>0$ in the strong-coupling regime, Figs. \ref{fig6}(d1)-(d3) indicate that the molecular spectrum still remains even at $T/T_{\rm c}^0=2.5$, because of a large binding energy. In this case, thermal effects on the atom-molecule coupling are dominated by thermal effect (1) in the above discussion. Indeed, in Figs. \ref{fig6}(c1)-(c3), while the molecular peak line gradually becomes broad with increasing the temperature, the existence of this coupling phenomenon itself can still be confirmed in $A_{\rm F}({\bm p},\omega)$  even at $T/T_{\rm c}^0=2.5$ (panel (c3)).
\par
Figure \ref{fig6} indicates that, when we use the Fermi SW to examine the crossover from the medium-assisted (many-body) bound state in the weak-coupling regime to the two-body bound state in the strong-coupling regime in a highly population-imbalanced Bose-Fermi mixture, we need to set the temperature near $T_{\rm c}$, in order to observe the former bound state.
\par
\begin{figure}[t]
\includegraphics[width=10cm]{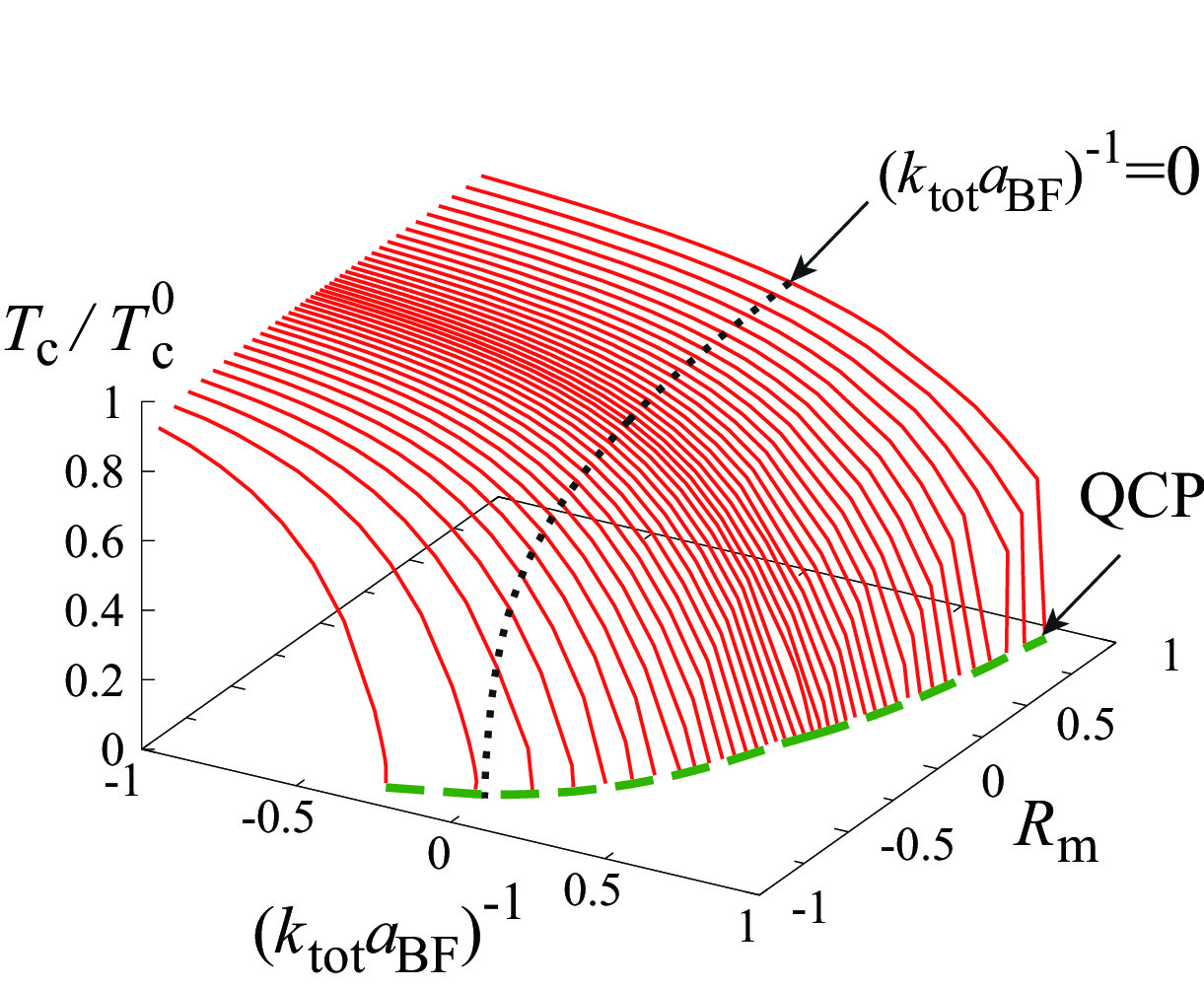}
\caption{Calculated BEC phase transition temperature $T_{\rm c}$ in a Bose-Fermi mixture with mass imbalance. The parameter $R_{\rm m}$ is defined in Eq. (\ref{eq.4.1}): $R_{\rm m}>0$ ($R_{\rm m}<0$) corresponds to the case of $m_{\rm F}>m_{\rm B}$ ($m_{\rm F}<m_{\rm B}$). The dashed line is the quantum critical point (QCP) at which $T_{\rm c}$ vanishes. $T_{\rm c}^0$ is the BEC phase transition temperature in an ideal Bose gas given in Eq. (\ref{eq.3.1}).
}
\label{fig7}
\end{figure}
\par
\begin{figure}
\includegraphics[width=10cm]{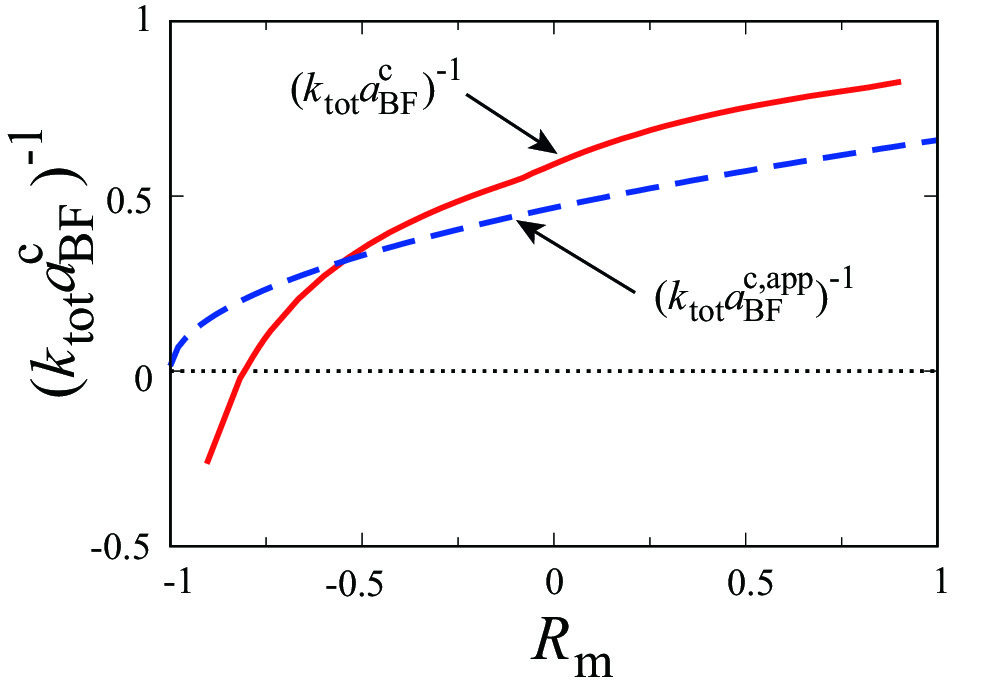}
\caption{Critical interaction strength $(k_{\rm tot}a_{\rm BF}^{\rm c})^{-1}$, which is defined as the interaction strength at which $T_{\rm c}$ vanishes, in a mass-imbalanced Bose-Fermi mixture. The dashed line shows the approximate result $(k_{\rm tot}a_{\rm BF}^{\rm c,app})^{-1}$ in Eq. (\ref{eq.4.3}). 
}
\label{fig8}
\end{figure}
\par
\section{Single-particle properties of mass-imbalanced Bose-Fermi mixture}
\par
Figure \ref{fig7} shows the BEC phase transition temperature $T_{\rm c}$ and effects of mass imbalance in a Bose-Fermi mixture, where mass difference is parametrized by
\begin{equation}
R_{\rm m}={m_{\rm F}-m_{\rm B} \over m_{\rm F}+m_{\rm B}}.
\label{eq.4.1}
\end{equation}
In this figure, $T_{\rm c}$ gradually decreases from the ideal Bose-gas result $T_{\rm c}^0$ in Eq. (\ref{eq.3.1}), with increasing the Bose-Fermi interaction strength, to eventually vanish at a certain interaction strength (QCP). Although this phenomenon has already been known in the mass balanced case\cite{cTMA1,iTMA1}, Fig. \ref{fig7} indicates that this suppression effect becomes more (less) remarkable when $m_{\rm F}<m_{\rm B}$ ($m_{\rm F}>m_{\rm B}$). To clearly show this, we separately plot in Fig. \ref{fig8} the critical interaction strength ($\equiv (k_{\rm tot}a_{\rm BF}^{\rm c})^{-1}$) at which $T_{\rm c}$ vanishes.
\par
The overall behavior of $(k_{\rm tot}a_{\rm BF}^{\rm c})^{-1}$ shown in Fig. \ref{fig8} may be understood as a result of competition between (1) the Bose-Einstein condensation of unpaired bosons, and (2) the formation of Bose-Fermi hetero-pairs: When $(k_{\rm tot}a_{\rm BF})^{-1}>0$, the two-body Bose-Fermi bound state can be formed, with the binding energy,
\begin{equation}
E_{\rm bind}^{\rm 2b}={1 \over ma_{\rm BF}^2},
\label{eq.4.2}
\end{equation}
where $m$ is given below Eq. (\ref{eq.2}). In the strong-coupling regime where most bosons form Bose-Fermi molecules, when the bare BEC transition temperature $T_{\rm c}^0$ is much lower than the characteristic temperature $T_{\rm bind}^{\rm 2b}\sim E_{\rm bind}^{\rm 2b}$, the BEC instability would no longer occur, because of the absence of unpaired bosons. Thus, the QCP is roughly estimated from the condition $E_{\rm bind}^{\rm 2b}\sim T_{\rm c}^0$, which approximately gives the critical interaction strength,
\begin{align}
(k_{\rm tot}a_{\rm BF}^{{\rm c,app}})^{-1}
&=
\left(
{4 \over 3\sqrt{\pi}\zeta(3/2)}
\right)^{1/3}
\frac{1}{\sqrt{1+m_{\rm B}/m_{\rm F}}}
\nonumber
\\
&=
0.66\frac{1}{\sqrt{1+m_{\rm B}/m_{\rm F}}}.
\label{eq.4.3}
\end{align}
Figure \ref{fig8} shows that Eq. (\ref{eq.4.3}) is consistent with the overall $R_{\rm m}$-dependence of $(k_{\rm tot}a_{\rm BF}^{\rm c})^{-1}$.
\par
\begin{figure}
\includegraphics[width=10cm]{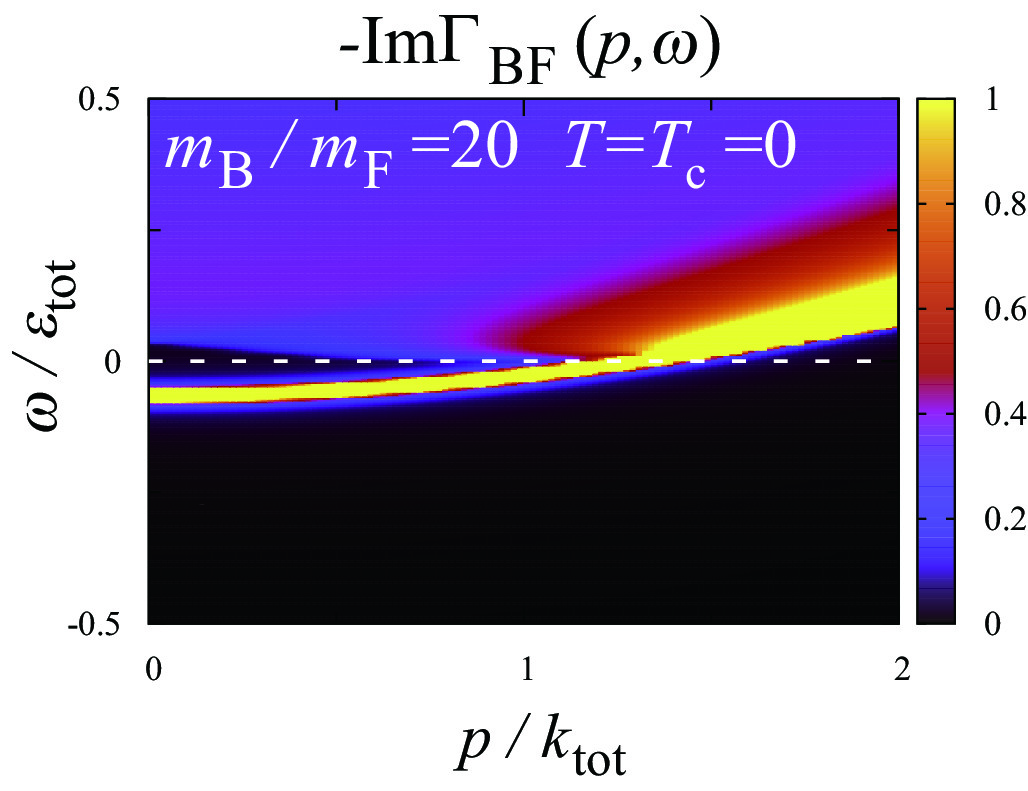}
\caption{Spectral intensity $-{\rm Im}[\Gamma_{\rm BF}(q,i\omega_{\rm F}\rightarrow\omega_+)]$ of the Bose-Fermi scattering matrix in a highly mass-imbalanced Bose-Fermi mixture ($m_{\rm B}/m_{\rm F}=20$, or $R_{\rm m}\simeq -0.9$). We take the interaction strength $(k_{\rm tot}a_{\rm BF})^{-1}$ to be equal to the critical value $(k_{\rm tot}a_{\rm BF}^c)^{-1}=-0.25$ at which $T_{\rm c}$ vanishes.}
\label{fig9}
\end{figure}

Because the above simple discussion assumes the two-body bound state, $(k_{\rm tot}a_{\rm BF}^{\rm c,app})^{-1}$ in Eq. (\ref{eq.4.3}) must be positive. However, a Bose-Fermi bound state actually exists also in the weak-coupling region $(k_{\rm tot}a_{\rm BF})^{-1}\le 0$, as a result of the medium effects. For example, we show in Fig. \ref{fig9} the spectrum $-{\rm Im}[\Gamma_{\rm BF}({\bm p},i\omega_{\rm F}\to\omega_+)]$ of Bose-Fermi scattering matrix in the highly mass-imbalanced case ($m_{\rm B}/m_{\rm F}=20\gg 1$) at the critical interaction strength $(k_{\rm tot}a_{\rm BF}^{\rm c})^{-1}=-0.25<0$ ($T=0$). In this figure, an isolated molecular branch is seen below the continuum, as in the population-imbalanced case (see Figs. \ref{fig4}-\ref{fig6}). This many-body bound state in the weak-coupling region ($(k_{\rm tot}a_{\rm BF})^{-1}\le 0$) naturally explains why the critical coupling $(k_{\rm tot}a_{\rm BF}^{\rm c})^{-1}$ can be negative in Fig. \ref{fig8}. 
\par
\par
\begin{figure}
\includegraphics[width=10cm]{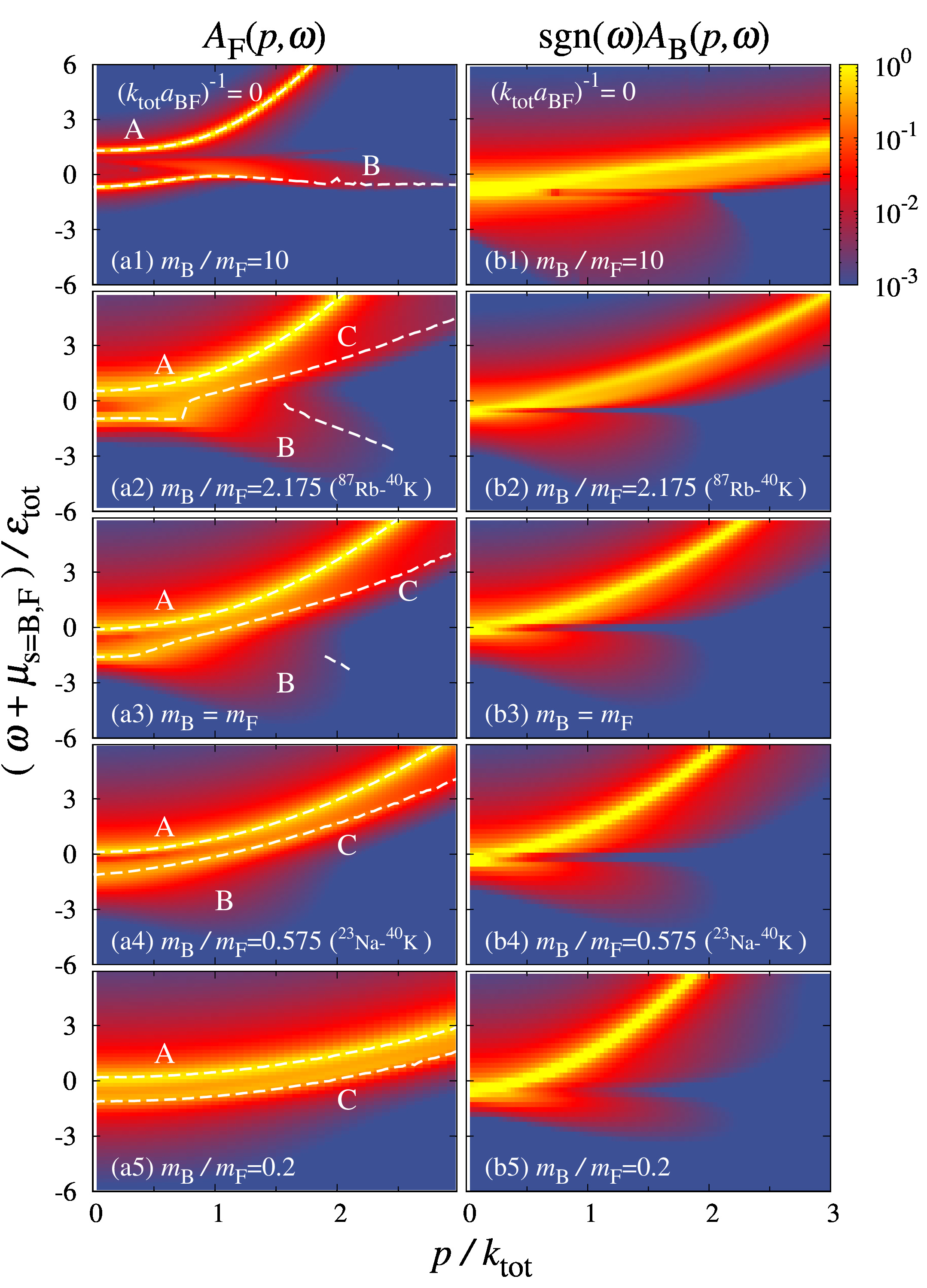}
\caption{Single-particle spectral weight in a unitary Bose-Fermi mixture with mass imbalance. We take $T=T_{\rm c}$, and $N_{\rm B}=N_{\rm F}$. The left and right panels show $A_{\rm F}(\bm{p},\omega)$ and $A_{\rm B}(\bm{p},\omega)$, respectively. The second panels from the top (bottom) show the case of a $^{87}$Rb-$^{40}$K ($^{23}$Na-$^{40}$K) mixture. For eye-guide, we plot the peak positions in the left figures (dashed lines).
}
\label{fig10}
\end{figure}
\par
Figure \ref{fig10} shows SWs $A_{\rm s=B,F}(\bm{p},\omega)$ in a mass-imbalanced unitary Bose-Fermi mixture at $T_{\rm c}$. In the Fermi SW (left panels), starting from the mass-balanced case (panel (a3)), we find that effects of mass difference are different between the cases of $m_{\rm B}/m_{\rm F}>1$ and $m_{\rm B}/m_{\rm F}<1$: Among the two sharp peaks ((A) and (C)) and a broad peak (B) in panel (a3), the peak (C), coming from the atom-molecule coupling, gradually disappears with increasing the ratio $m_{\rm B}/m_{\rm F}>1$ (see Fig. \ref{fig10}(a3)$\to$(a1)). When $m_{\rm B}/m_{\rm F}$ decreases from unity, on the other hand, Figs. \ref{fig10}(a3)$\to$(a5) show that the broad peak (B), originating from the Fermi-Bose coupling, gradually disappears. 
\par
\begin{figure}
\includegraphics[width=10cm]{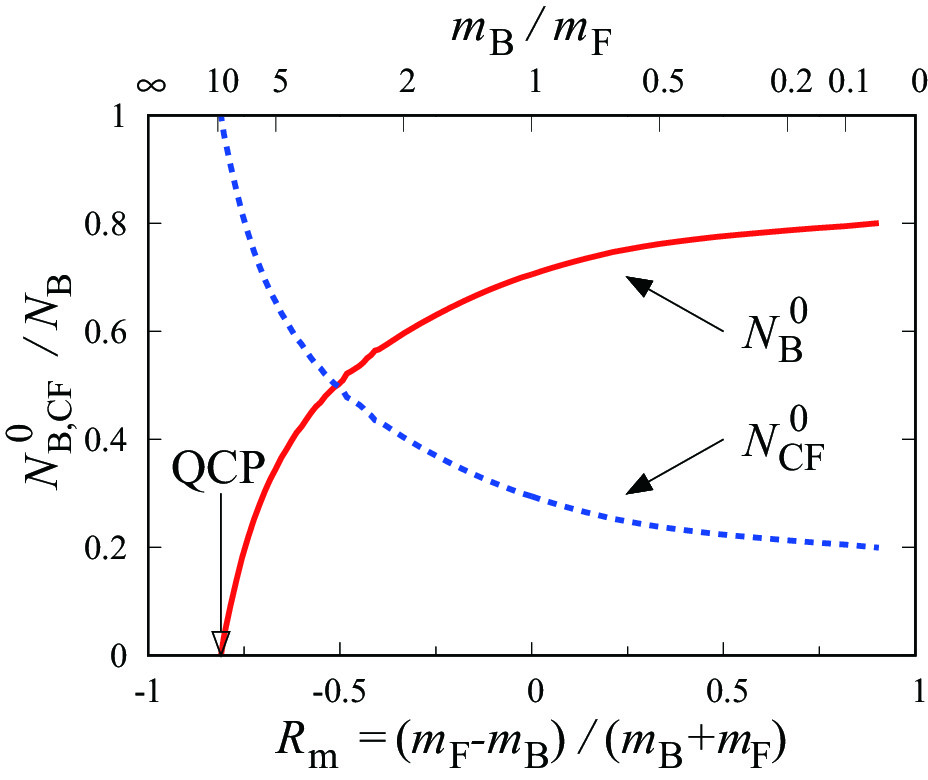}
\caption{The number $N_{\rm B}^0=\sum_{\bm p}({\tilde \xi}_{\bm p}^{\rm B})$ of unpaired Bose atoms, as well as the number $N_{\rm CF}^0\equiv N_{\rm B}-N_{\rm B}^0$ of the Bose-Fermi molecules in a unitary Bose-Fermi mixture at $T_{\rm c}$, as functions of mass imbalance parameter $R_{\rm m}$ in Eq. (\ref{eq.4.1}). Since the direct evaluation of $N_{\rm CF}=\sum_{\bm p}f(\xi_{\bm p}^{\rm CF})$ in Eq. (\ref{eq.3.3}) is difficult, we approximately use $N_{\rm CF}^0$ for $N_{\rm CF}$ in our discussions.
}
\label{fig11}
\end{figure}

To understand these in a simple manner, we conveniently plot in Fig. \ref{fig11} $N_{\rm B}^0=\sum_{\bm p}n_{\rm B}({\tilde \xi}_{\bm p}^{\rm B})$ and $N_{\rm CF}^0\equiv N_{\rm B}-N_{\rm B}^0~(\sim N_{\rm CF}=\sum_{\bm p}f(\xi_{\bm p}^{\rm CF}))$ in a unitary Bose-Fermi mixture at $T_{\rm c}$. Noting that these quantities are directly related to the atom-molecule coupling and Fermi-Bose coupling, respectively (see Eq. (\ref{eq.3.3})), we find from Fig. \ref{fig11} that the former (latter) coupling phenomenon becomes dominant when $R_{\rm m}\to 1$ ($R_{\rm m}\to -1$). In addition, the momentum dependence of the Bose kinetic energy ${\tilde \xi}_{\bm p}^{\rm B}$ becomes weak with increasing $m_{\rm B}$, so that the broadening by the angular integration in the last term the denominator in Eq. (\ref{eq.3.3}) is suppressed. Because of this, the broad peak (B) in the mass balanced case in Fig. \ref{fig10}(a3) gradually becomes sharp, as one moves from panels (a3) to (a1). 
\par
\par
Compared to the Fermi SW, the Bose SW $A_{\rm B}({\bm p},\omega)$ is not so sensitive to the mass imbalance, as shown in the right panels in Fig. \ref{fig10}, which is simply due to the two angular integrations in the denominator in Eq. (\ref{eq.3.6}). Thus, as in the population-imbalanced case, the Fermi SW $A_{\rm F}({\bm p},\omega)$ is more suitable for the study of strong-coupling corrections to single-particle excitations in a mass-imbalanced Bose-Fermi mixture.
\par
\begin{figure}
\includegraphics[width=10cm]{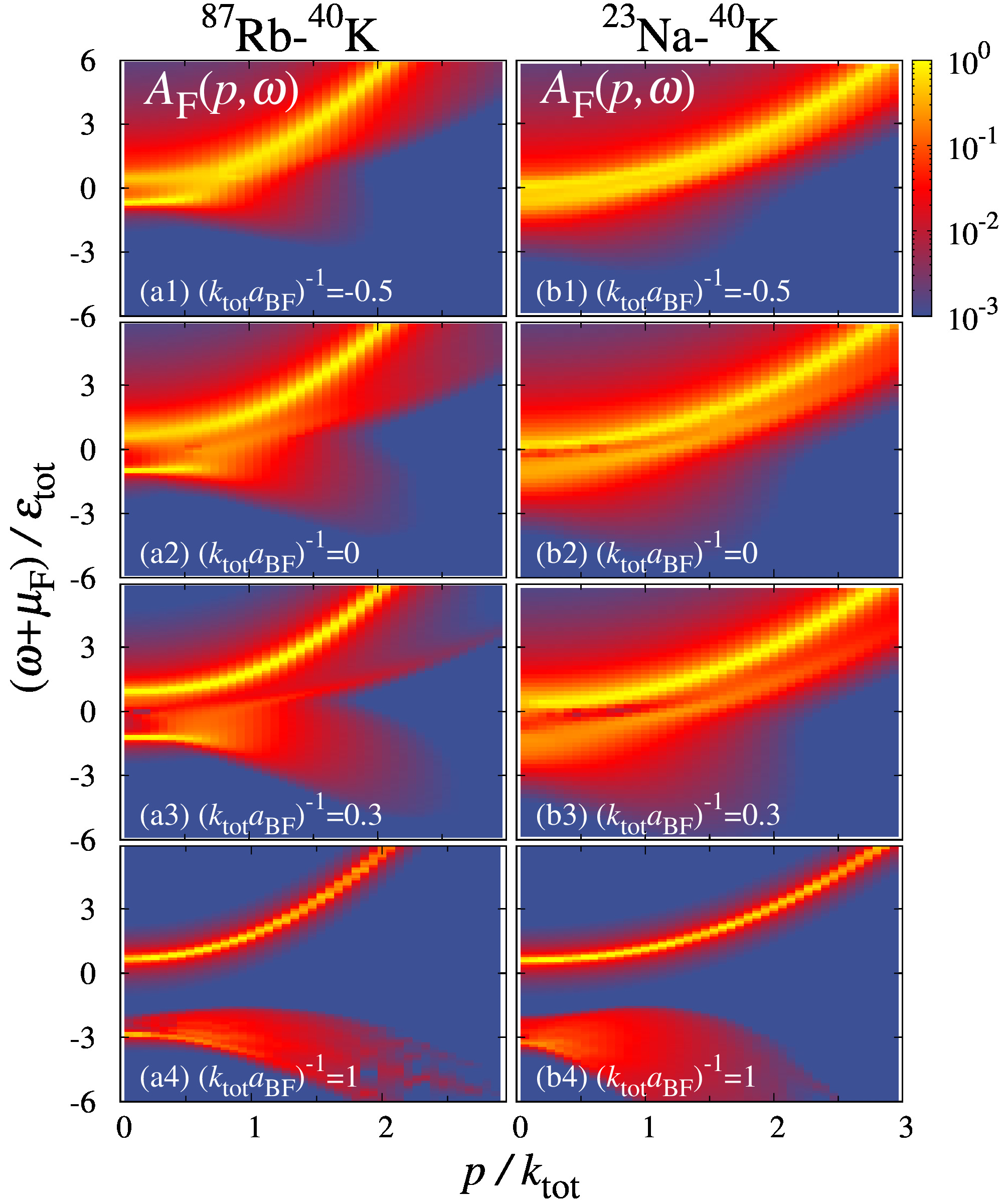}
\caption{Calculated Fermi SW $A_{\rm F}(\bm{p},\omega)$ in a $^{87}$Rb-$^{40}$K mixture (left panels), as well as a $^{23}$Na-$^{40}$K mixture (right panels). We set $T=T_{\rm c}$ for the upper three cases, and set $T=0.01T_{\rm c}^0$ for the lowest case. We note that QCP is at $(k_{\rm tot}a_{\rm BF}^{\rm c})^{-1}=0.43$ for $^{87}$Rb-$^{40}$K mixture, and $(k_{\rm tot}a_{\rm BF}^{\rm c})^{-1}=0.7$ for $^{23}$Na-$^{40}$K. Thus, there is no BEC phase transition when $(k_{\rm tot}a_{\rm BF})^{-1}=1$ in panels (a4) and (b4).
}
\label{fig12}
\end{figure}
\par
\begin{figure}
\includegraphics[width=10cm]{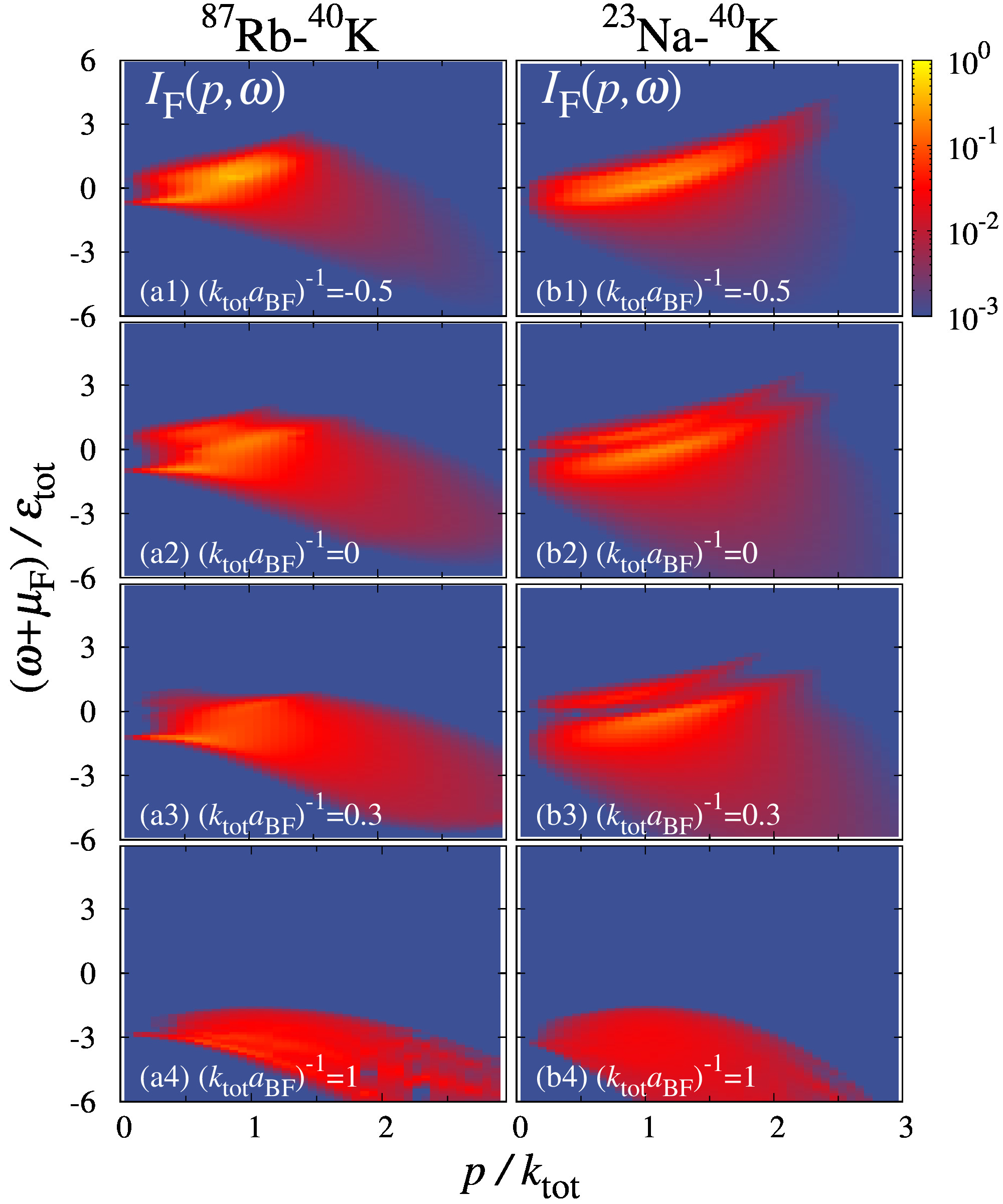}
\caption{Calculated Fermi photoemission spectrum (PES) $I_{\rm F}(p,\omega)$. Left panels: $^{87}$Rb-$^{40}$K mixture. Right panels: $^{23}$Na-$^{40}$K mixture. The parameters are same as those in Fig. \ref{fig12}. The spectral intensity is normalized by $4\pi t_{\rm F}^2m$. This normalization is also used in Figs. \ref{fig14} and \ref{fig15}.
}
\label{fig13}
\end{figure}
\par
\begin{figure}
\includegraphics[width=10cm]{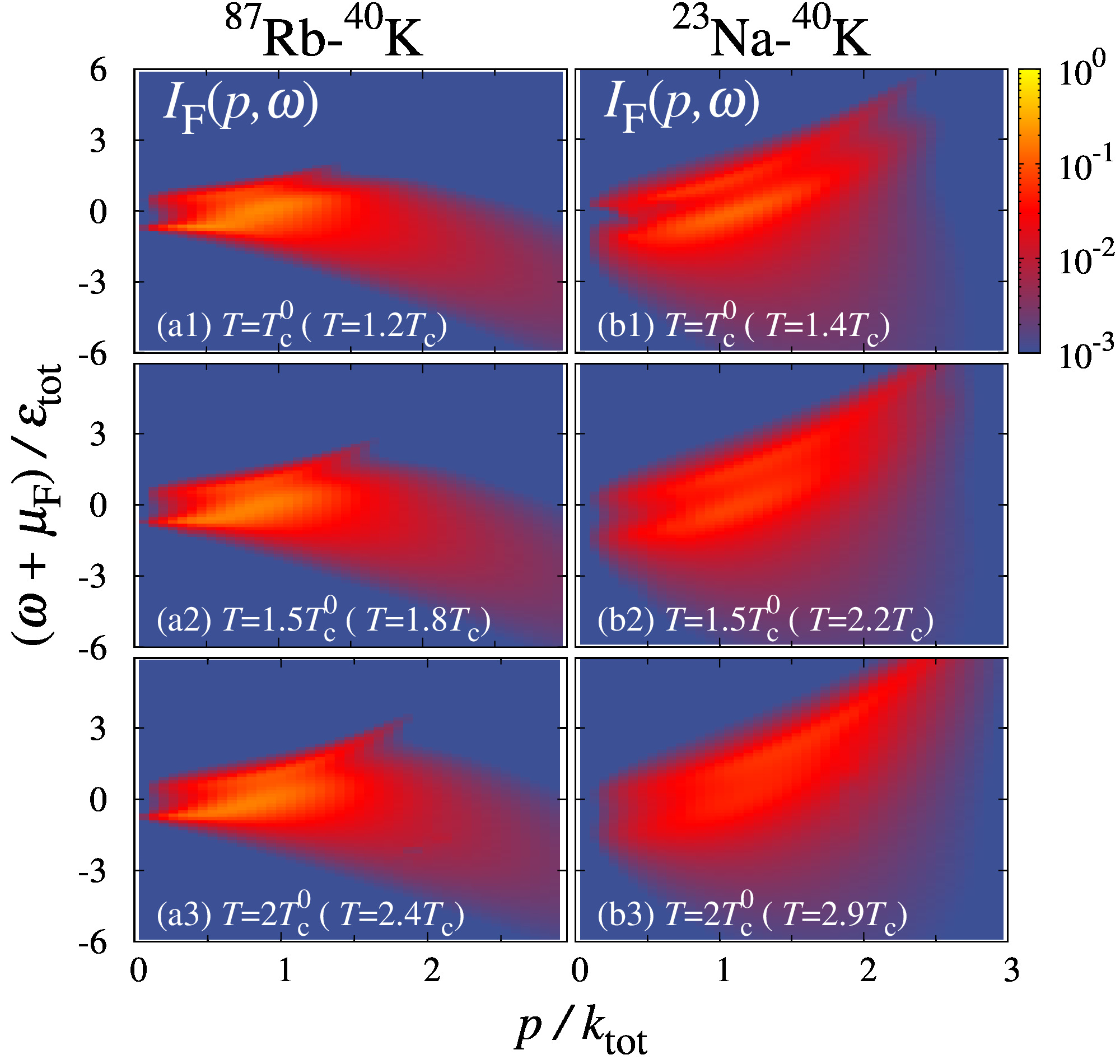}
\caption{Fermi PES $I_{\rm F}({\bm p},\omega)$ in the unitarity limit above $T_{\rm c}$. Left panels: $^{87}$Rb-$^{40}$K mixture ($T_{\rm c}=0.695T_{\rm c}^0$). Right panels: $^{23}$Na-$^{40}$K mixture ($T_{\rm c}=0.828T_{\rm c}^0$). The results at $T_{\rm c}$ are shown in Figs. \ref{fig13}(a2) and (b2). 
}
\label{fig14}
\end{figure}
\par
\begin{figure}
\includegraphics[width=10cm]{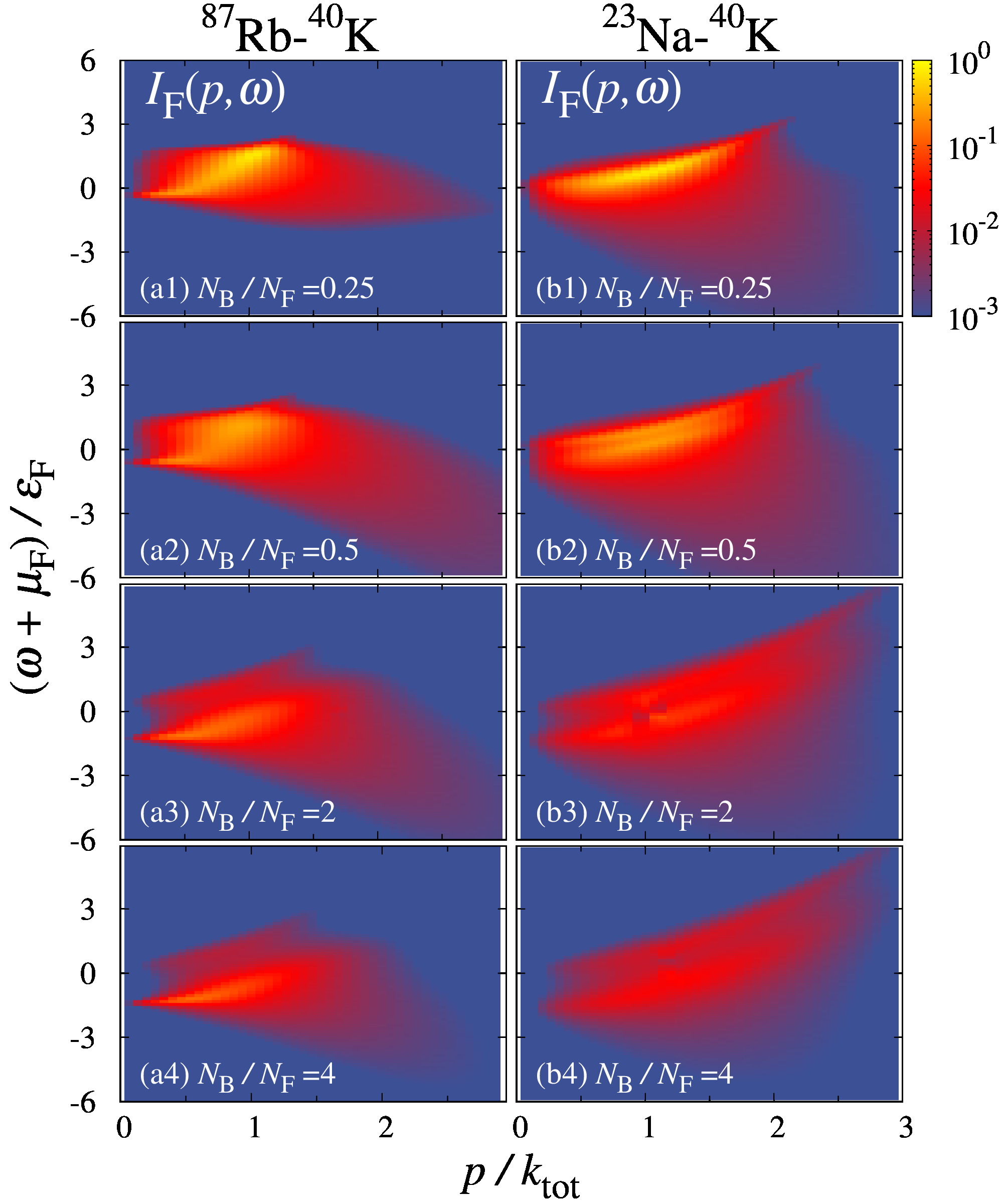}
\caption{Effects of population imbalance on the Fermi PES $I_{\rm F}(p,\omega)$ in a unitary Bose-Fermi mixture at $T=1.5T_{\rm c}$. Left panels: $^{87}$Rb-$^{40}$K mixture. Right panels: $^{23}$Na-$^{40}$K mixture. For the population-balanced results, see Figs. \ref{fig13}(a2) and (b2).
}
\label{fig15}
\end{figure}
\par
\section{Examples: $^{87}$Rb-$^{40}$K and $^{23}$Na-$^{40}$K mixtures}
\par
Figures \ref{fig10}(a2) and (a4) show the cases of existing $^{87}$Rb-$^{40}$K and $^{23}$Na-$^{40}$K mixtures, respectively. These figures predict that the Fermi SW still exhibits the three-peak structure in the $^{87}$Rb-$^{40}$K case. Although the broad peak (B) is suppressed in a $^{23}$Na-$^{40}$K mixture, we can still see the atom-molecule coupling in this mixture.
\par
We emphasize that these many-body coupling phenomena still remain to some extent, away from the unitarity limit, as shown in Fig. \ref{fig12}. In both the mixtures, the atom-molecule coupling is found to remain with {\it decreasing} the interaction strength (see the upper two panels in Fig. \ref{fig12}). This is simply due to the increase of $N_{\rm B}^0$ and the decrease of $N_{\rm CF}$ in the denominator in Eq. (\ref{eq.3.3}). On the other hand, as one {\it increases} the interaction strength (see the lowest two panels in Fig. \ref{fig12}), the Fermi-Bose coupling becomes important, reflecting the decrease of $N_{\rm B}^0$ and the increase of $N_{\rm CF}$. In Figs. \ref{fig12}(a4) and (b4), the Fermi-Bose coupling is only seen.
\par
Figure \ref{fig13} shows the Fermi photoemission spectrum (PES) $I_{\rm F}({\bm p},\omega)$ in a Bose-Fermi mixture, where the parameters in each panel are the same as those in the corresponding panel in Fig. \ref{fig12}. Comparing Fig. \ref{fig12} with Fig. \ref{fig13}, one finds that, although the spectral structure seen in the positive energy region of SW is suppressed by the Fermi distribution function in PES (see Eq. (\ref{eq.15})), it can still detect the downward broad spectral structure coming from the Fermi-Bose coupling associated with hetero-pairing fluctuations. 
\par
For the upward peak line along the molecular dispersion seen in the Fermi SW, since the thermal broadening of the Fermi distribution function $f(\omega)$ around $\omega=0$ weakens the suppression of the spectral intensity in the positive energy region in the Fermi PES, it gradually appears in PES with increasing the temperature, as shown in Fig. \ref{fig14}. Of course, this method to observe the molecular branch is not always valid, especially for the very high-temperature region where Bose-Fermi molecules thermally dissociate into unpaired atoms. However, since the molecular binding energy is large in the strong coupling regime, the temperature region where this idea works would be wide there.
\par
Figure \ref{fig15} shows effects of population imbalance on the Fermi PES $I_{\rm F}({\bm p},\omega)$. As expected from Fig. \ref{fig3}, the multiple peak structure gradually becomes obscure in $I_{\rm F}({\bm p},\omega)$ with decreasing the ratio $N_{\rm B}/N_{\rm F}$ from unity. Simply interpreting this result as the local PES at various spatial positions in a trapped Bose-Fermi mixture, one can imagine that the detailed spectral structure is smeared out after the spatial average of the spectrum, when the contribution from the spatial region where the Bose atomic density $\rho_{\rm B}({\bm r})$ is much smaller than the Fermi atomic density $\rho_{\rm F}({\bm r})$ is dominant. To avoid this as possible as we can, it would be a good idea to detect spectra, avoiding the spatial region where $\rho_{\rm B}({\bm r})\ll \rho_{\rm F}({\bm r})$. For this purpose, the local photoemission-type experiment developed by JILA group\cite{PES_exp3} would be useful. As an alternative way, a box trap\cite{box_Fermi,box_Bose1,box_Bose2} may also be promising, because an almost uniform gas is realized there. 
\par
\par
\section{Summary}
\par
To summarize, we have discussed single-particle excitations and effects of mass and population imbalances in a Bose-Fermi mixture. Including hetero-pairing fluctuations associated with an attractive Bose-Fermi interaction within the framework of the improved $T$-matrix approximation developed by two of the authors, we calculated the single-particle spectral weight (SW), as well as the photoemission spectrum (PES), in the normal state above the BEC phase transition temperature $T_{\rm c}$.
\par
In the mass- and population-balanced case ($m_{\rm B}=m_{\rm F}$ and $N_{\rm B}=N_{\rm F}$), it is known that strong hetero-pairing fluctuations cause couplings between atomic excitations and composite molecular excitations (atom-molecule coupling), as well as between Fermi atomic excitations and Bose atomic excitations (Fermi-Bose coupling). These many-body phenomena bring about additional two spectral peaks in the Fermi SW. Together with the ordinary spectral peak along the single-particle Fermi dispersion, the resulting Fermi SW exhibits a three-peak structure.
\par
In the presence of population imbalance, we showed that, when $N_{\rm B}/N_{\rm F}\ll 1$, the both atom-molecule and Fermi-Bose coupling phenomena become weak, so that the Fermi SW becomes close to that in a free Fermi gas. When $N_{\rm B}/N_{\rm F}\gg 1$, on the other hand, the former coupling remains to exist, leading to a two-peak structure in the Fermi SW. This difference comes from the fact that, while the atom-molecule coupling constant is dominated by the number of unpaired Bose atoms, the Fermi-Bose coupling constant is deeply related to the number $N_{\rm CF}$ of Bose-Fermi molecules ($N_{\rm CF}\le\min(N_{\rm B},N_{\rm F})$): The both coupling constants thus become small when $N_{\rm B}/N_{\rm F}\ll 1$. On the other hand, the former coupling remains non-zero even when $N_{\rm B}/N_{\rm F}\gg 1$. 
\par
We have also examined how mass difference between a Fermi atom ($m_{\rm F}$) and a Bose atom ($m_{\rm B}$) modifies many-body corrections to single-particle excitations. In both the limits $m_{\rm B}/m_{\rm F}\ll 1$ and $m_{\rm B}/m_{\rm F}\gg 1$, we found that the Fermi SW exhibits, not a three-peak, but a two-peak structure; however, their physical meanings are different. When $m_{\rm B}/m_{\rm F}\ll 1$, the Fermi-Bose coupling causes the second peak line in addition to the ordinary peak line along the free-particle dispersion. In the opposite limit, the additional peak comes from the atom-molecule coupling. This is because the strengths of these couplings differently depend on the ratio $m_{\rm B}/m_{\rm F}$. 
\par
When one goes away from these limiting cases, the Fermi SW exhibits the three-peak structure as in the mass-balanced case. We explicitly confirmed this in the cases of a mass-imbalanced $^{87}$Rb-$^{40}$K ($m_{\rm B}>m_{\rm F}$) and a $^{23}$Na-$^{40}$K ($m_{\rm B}<m_{\rm F}$) mixtures. We also pointed out that these many-body coupling phenomena may be observed by the photoemission-type experiment, by explicitly evaluating the photoemission spectra for these realistic examples.
\par
In this paper, we have treated a uniform Bose-Fermi mixture, for simplicity. In a real trapped mixture in a harmonic potential, we expect that the Fermi and Bose atoms have different density profiles, leading to local population imbalance. Although this inhomogeneous effect has only partially been examined in this paper, by considering the population-imbalanced case, to fully understand strong-coupling properties of a trapped Bose-Fermi mixture, it would be necessary to explicitly treat the trapped geometry. Besides this, we have also ignored an interaction between Bose atoms, which would be crucial for the stability of this system. These problems remain as our future challenges. Since the atom-molecule and Fermi-Bose couplings are characteristic many-body phenomena in a Bose-Fermi mixture with a hereto-pairing interaction, our results would contribute to further understanding of strong-coupling properties of this novel quantum many-body system.

\par
\par
\begin{acknowledgments}
We thank D. Kagamihara and R. Sato for discussions. This work was supported by the KiPAS project at Keio University. Y.O. was supported by a Grant-in-aid for Scientific Research from MEXT and JSPS in Japan (No.JP18K11345, No.JP18H05406, and No.JP19K03689).
\end{acknowledgments}
\par
\par
\appendix
\par
\section{Derivation of Eq. (\ref{eq.3.3})}
\par
We approximate the Bose-Fermi scattering matrix $\Gamma_{\rm BF}({\bm q},i\omega_{\rm F})$ in Eq. (\ref{eq.10}) to the composite molecular propagator,
\begin{equation}
\Gamma_{\rm BF}({\bm q},i\omega_{\rm F})\simeq
{Z \over i\omega_{\rm F}-\xi_{\bm{q}}^{\rm CF}},
\label{eq.A1}
\end{equation}
where $Z$ is a positive constant\cite{iTMA1}, and the molecular dispersion $\xi_{\bm p}^{\rm CF}$ is given below Eq. (\ref{eq.3.2}). Although Eq. (\ref{eq.A1}) is, strictly speaking, only justified in the strong-coupling limit, this approximate expression is still useful to grasp the essence of strong-coupling corrections to single-particle excitations. Substituting Eq. (\ref{eq.A1}) into the self-energy $\Sigma_{\rm F}({\bm p},i\omega_{\rm F})$ in Eq. (\ref{eq.8}), we have, after carrying out the $\omega'_{\rm F}$-summation, 
\begin{eqnarray}
\Sigma_{\rm F}({\bm p},i\omega_{\rm F})
&=&
Z\sum_{\bm q}
\left[
{n_{\rm B}({\tilde \xi}_{\bm q}^{\rm B}) 
\over 
i\omega_{\rm F}-\xi^{\rm CF}_{{\bm p}-{\bm q}}
+{\tilde \xi}_{\bm q}^{\rm B}}
+
{f(\xi_{\bm q}^{\rm CF}) \over
i\omega_{\rm F}+{\tilde \xi}_{{\bm p}-{\bm q}}^{\rm B}-\xi_{\bm q}^{\rm CF}}
\right]
\nonumber
\\
&\simeq&
{Z{\tilde N}_{\rm B}^0 \over i\omega_{\rm F} -\xi_{\bm p}^{\rm CF}}
+
\left\langle
{ZN_{\rm CF} \over i\omega_{\rm F}+{\tilde \xi}_{{\bm k}_{\rm CF}-{\bm p}}^{\rm B}}
\right
\rangle_{{\bm k}_{\rm CF}}
.
\label{eq.A2}
\end{eqnarray}
Here, the definitions of the parameters in Eq. (\ref{eq.A2}) are explained in the text. In obtaining the first term in the last line, we have set ${\bm q}=0$ in the denominator by using that the Bose distribution function $n_{\rm B}({\tilde \xi}_{\bm q}^{\rm B})$ diverges in the low momentum limit at $T_{\rm c}$. For the last term in the last line in Eq. (\ref{eq.A2}), we have approximated ${\bm q}$ in the denominator to ${\bm k}_{\rm CF}$, by noting that the region near the Fermi surface of the composite Fermi molecules is important. Substitution of Eq. (\ref{eq.A2}) into Eq. (\ref{eq.3}) gives Eq. (\ref{eq.3.3}).
\par


\begin{thebibliography}{99}
\bibitem{comment4} Although a two-body bound state vanishes at the unitary,  medium effects gives a bound state at this interaction strength, as will be explained later.
\end{thebibliography}
\end{document}